\documentclass[aps,a4paper,twocolumn,pra,superscriptaddress,longbibliography]{revtex4-1}
\usepackage[english]{babel}
\usepackage[T1]{fontenc}
\usepackage{amssymb,amsthm,mathrsfs,amscd,amsmath,braket}
\usepackage[breaklinks=true,colorlinks=true,linkcolor=blue,urlcolor=blue,citecolor=blue]{hyperref}
\usepackage{graphicx}
\usepackage[dvipsnames]{xcolor}

\begin{document}

\title{The quantized Hall conductance of a single atomic wire: \\ A proposal based on synthetic dimensions}

\author{G. Salerno}
\affiliation{Center for Nonlinear Phenomena and Complex Systems, Universit\'{e} Libre de Bruxelles, CP 231, Campus Plaine, B-1050 Brussels, Belgium}
\author{H. M. Price}
\affiliation{School of Physics and Astronomy, University of Birmingham, Edgbaston, Birmingham B15 2TT, United Kingdom}
\author{M. Lebrat}
\affiliation{Institute for Quantum Electronics, ETH Zurich, 8093 Zurich, Switzerland}
\author{S. H\"ausler}
\affiliation{Institute for Quantum Electronics, ETH Zurich, 8093 Zurich, Switzerland}
\author{T. Esslinger}
\affiliation{Institute for Quantum Electronics, ETH Zurich, 8093 Zurich, Switzerland}
\author{L. Corman}
\affiliation{Institute for Quantum Electronics, ETH Zurich, 8093 Zurich, Switzerland}
\author{J.-P. Brantut}
\affiliation{Institute of Physics, EPFL, 1015 Lausanne, Switzerland}
\author{N. Goldman}
\affiliation{Center for Nonlinear Phenomena and Complex Systems, Universit\'{e} Libre de Bruxelles, CP 231, Campus Plaine, B-1050 Brussels, Belgium}

\date{\today}

\begin{abstract}

We propose a method by which the quantization of the Hall conductance can be directly measured in the transport of a one-dimensional atomic gas. Our approach builds on two main ingredients: (1) a constriction optical potential, which generates a mesoscopic channel connected to two reservoirs, and (2) a time-periodic modulation of the channel, specifically designed to generate motion along an additional synthetic dimension. This fictitious dimension is spanned by the harmonic-oscillator modes associated with the tightly-confined channel, and hence, the corresponding ``lattice sites'' are intimately related to the energy of the system. We analyze the quantum transport properties of this hybrid two-dimensional system, highlighting the appealing features offered by the synthetic dimension. In particular, we demonstrate how the energetic nature of the synthetic dimension, combined with the quasi-energy spectrum of the periodically-driven channel, allows for the direct and unambiguous observation of the quantized Hall effect in a two-reservoir geometry. Our work illustrates how topological properties of matter can be accessed in a minimal one-dimensional setup, with direct and practical experimental consequences. 
\end{abstract}

\maketitle

\section{Introduction}
\label{Sec:Intro}

Electronic transport in solids plays a fundamental role in our exploration of matter, and it constitutes the basis for innumerable device applications. In fact, the need for smaller and more efficient hardware has naturally led to the development of mesoscopic devices, where the quantum nature of the electron gas becomes relevant~\cite{Imry_book}. One of the most prominent examples of such quantum phenomena is the quantization of the electrical conductance in mesoscopic channels, which stems from the existence of discrete transverse modes~\cite{vanWees, Wharam, Krans,Landauer, Buttiker}. In the 1980's, studies of the Hall conductance in two-dimensional electron gases subjected to high magnetic fields revealed the quantized Hall effect~\cite{vonKlitzing, Goerbig, Yoshioka}, which was later related to the existence of topological invariants in the band structure~\cite{Thouless, Niu} and chiral edge modes~\cite{Laughlin, Halperin1982, macdonald1984edge, hatsugai1993chern}. Moreover, such Hall measurements subsequently revealed the fractional quantum Hall effect \cite{Tsui}, a first instance of a strongly-correlated topological phase~\cite{Laughlin1983, Haldane1983}. More recent transport experiments revealed the existence of topological insulators, such as those realizing the quantum spin Hall effect~\cite{KaneMele, KaneMele2, Bernevig, Konig}, as well as Dirac and Weyl semimetals~\cite{Hosur2013,Armitage}. In this regard, transport measurements are an important and well-established method for probing and studying the properties of quantum matter~\cite{Imry, Datta, Amico}.

In parallel to the exploration of new materials and devices, quantum-engineered systems have been developed in ultracold-atom laboratories in view of offering a novel perspective on transport in quantum matter~\cite{Bloch}. In these settings, non-equilibrium dynamics can be probed through different protocols~\cite{Krinner_Rev}, for instance, by suddenly releasing the atomic cloud in an optical lattice and imaging its expansion~\cite{Schneider2012,Scherg2018,Brown2018}, or by driving the cloud with an external (optical or magnetic) force~\cite{Beeler,Dahan1996,Jotzu,Aidelsburger,Anderson,Asteria}. A third approach consists in engineering a two-terminal geometry, i.e.~a mesoscopic channel for atoms connected to two reservoirs, using a constriction optical potential~\cite{Brantut}: this scheme, which reproduces the two-terminal configuration used in electronic transport experiments, allows for a direct evaluation of a neutral gas' conductance. Importantly, such a setting has demonstrated the quantized conductance of a one-dimensional atomic channel~\cite{Krinner_Nature}, which constitutes a good starting point to study the transport of strongly correlated matter thanks to the ability to tune the interaction strength~\cite{Husmann2015}. A particularly exciting perspective concerns the observation and characterization of fractional quantum Hall states in ultracold atomic gases~\cite{Cooper2008,Goldman2016,Cooper2018}.

Measuring the quantized Hall conductance of a two-dimensional (2D) ultracold Fermi gas, using the engineered-reservoir scheme of Refs.~\cite{Brantut,Krinner_Nature}, is definitely appealing. However, this challenging goal would a priori require the combination of two main ingredients: (1) the realization of a synthetic gauge field~\cite{Cooper2008,Dalibard,Goldman2014} to create a non-trivial topological band structure and reach the quantum-Hall regime~\cite{Goldman2016,Cooper2018}; and (b) in direct analogy with the multi-terminal devices (Hall-bar geometries) used in solid-state experiments~\cite{vonKlitzing, Goerbig, Yoshioka}, one would need to connect the 2D Fermi gas to several reservoirs. In this work, we propose that such an apparently complicated setting could in fact be readily designed starting from a \emph{single atomic channel}~\cite{Krinner_Nature}, by exploiting the concept of synthetic dimensions~\cite{Boada, Celi, Cooper, Gadway, Mancini, Stuhl, Meier, Livi, An, Price, Kolkowitz}. As we will show, the use of a synthetic dimension does not only simplify the implementation of the ``atomic Hall bar'', but it also allows for a direct read-out of the quantized Hall conductance associated with chiral edge modes using a simple \emph{two-reservoir} geometry. This scheme is universal in the sense that it could be applied to study the Hall conductance of a wide variety of atomic states (with potential applications to strongly-correlated states). Besides, we note that the main concepts developed in this proposal could also be used in other physical platforms; see, for instance, the recent proposal in Ref.~\cite{Bauer2018} to braid Majorana fermions in a single superconducting wire extended by a synthetic dimension; see also Refs.~\cite{Ozawa2016,Yuan_Review, Lustig, Lin2018} for synthetic dimensions in photonics. 

\begin{figure}[t]
\centering
\includegraphics[width=1 \columnwidth]{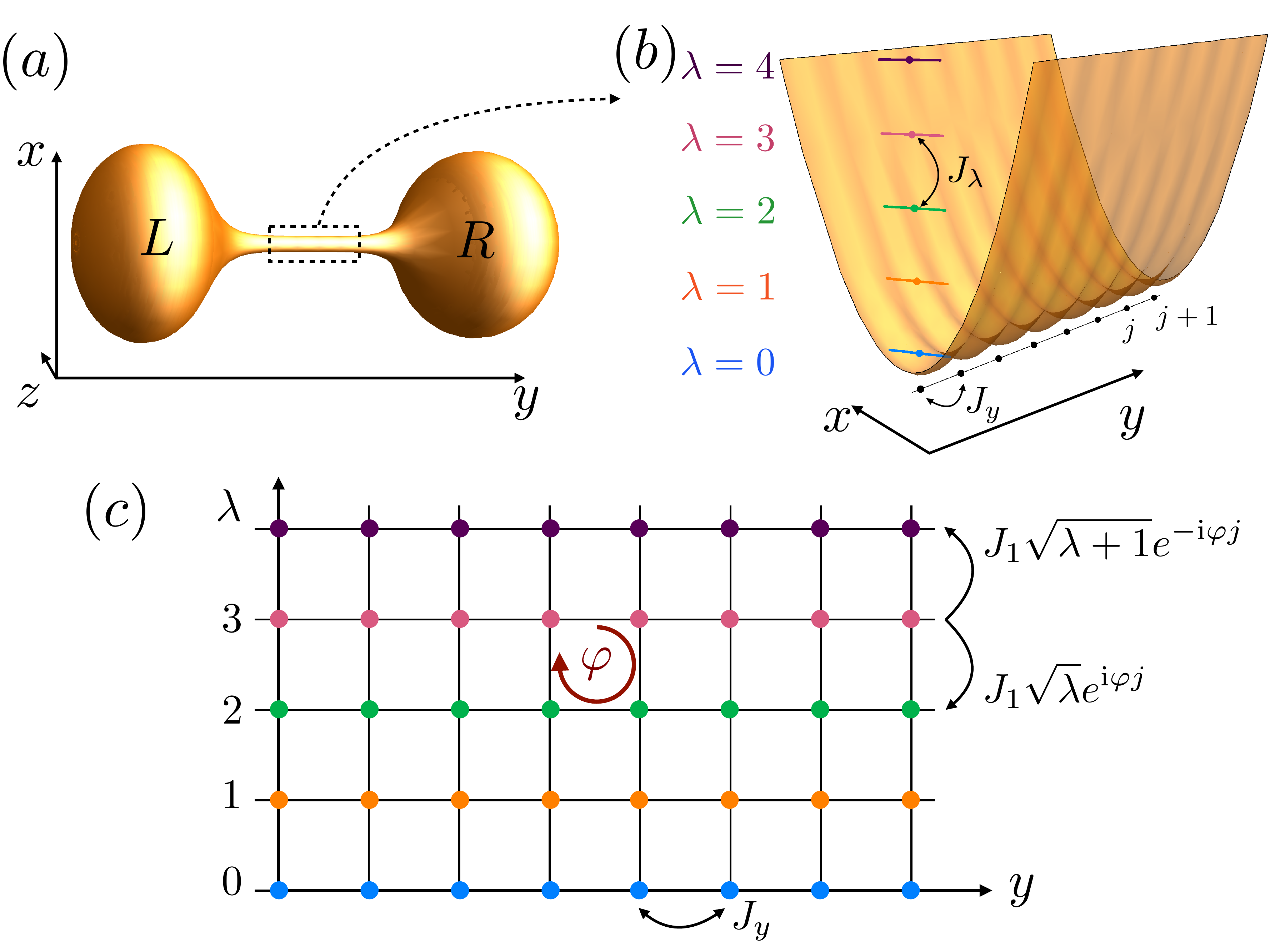}
\caption{(a) A sketch of the two-reservoir geometry~\cite{Krinner_Rev}, which is used as a basis for our shaken-channel model. A constriction potential is imprinted on a 3D Fermi gas so as to create a quantum wire, on top of which, a one-dimensional optical lattice is produced. The non-constricted parts of the system form the two atomic reservoirs at the left ($L$) and right ($R$) sides of the channel. (b) A close-up view on the channel region shows the harmonic confinement associated with the constriction potential, as well as the optical lattice aligned along the channel; the corresponding sites are indexed by $j$ and the hopping amplitude denoted $J_y$. The harmonic oscillator levels $\lambda$ are reinterpreted as lattice sites along a synthetic dimension. The potential is then periodically shaken, as proposed in Ref.~\cite{Price}, to generate hopping processes along the synthetic dimension $\lambda$; the corresponding hopping amplitude is denoted $J_\lambda\!=\!J_1\sqrt{\lambda}$. (c) For a suitable shaking protocol, the system is described by a 2D tight-binding model in the $y-\lambda$ plane, which includes the effects of a tunable artificial magnetic field; the corresponding flux per plaquette is denoted by $\varphi$.}
\label{Fig:TransportChannel}
\end{figure}

\subsection{The main approach and central results}\label{Sec:Approach}

 The aim of this work is to lay out a scheme by which a single atomic wire, connected to two engineered reservoirs [Fig.~\ref{Fig:TransportChannel}(a)], can be turned into an ``atomic Hall bar'': a 2D atomic system exhibiting the quantum Hall effect and designed so as to extract its (quantized) Hall conductance through transport. 

Our approach is based on the observation that the constriction potential used to generate the atomic channel~\cite{Brantut} defines a natural synthetic dimension. As illustrated in Fig.~\ref{Fig:TransportChannel}(b), the harmonic-oscillator levels associated with the tight confinement form a large set of discrete states, indexed by $\lambda\!\in\!\mathbb{N}$, which can be interpreted as fictitious ``lattice sites'' along a synthetic dimension. As shown in Ref.~\cite{Price}, motion can be induced along the $\lambda$ direction by shaking the channel in a time-periodic manner. Hence, within the central region of the constriction potential, atoms are allowed to move along the real direction defined by the channel (denoted ``$y$ direction'' in Fig.~\ref{Fig:TransportChannel} and hereafter), as well as along the synthetic dimension $\lambda$. We note that this construction essentially replaces the continuous transverse direction $x$ [Fig.~\ref{Fig:TransportChannel}(a)] by a discrete synthetic dimension $\lambda$ [Fig.~\ref{Fig:TransportChannel}(b)], and that the motion is inhibited along the third spatial direction ($z$); besides this, we will assume that the channel direction ($y$) can also be discretized upon adding an optical lattice, as recently implemented in Ref.~\cite{Lebrat}. As a final ingredient, we will assume that the phase of the modulation that generates motion along $\lambda$ can be made $y$-dependent, $\phi (y)$: as previously shown in Ref.~\cite{Price}, this can generate a uniform magnetic flux in the 2D lattice defined in the fictitious $\lambda-y$ plane [Fig.~\ref{Fig:TransportChannel}(c)]; see also Refs.~\cite{Kolovsky, Kolovsky_Comment, Bermudez2011,Goldman2015,Creffield2016}. In the following, we will consider that the reservoirs are not subjected to the modulation, and hence, that the corresponding regions do not include a synthetic magnetic field.

This work analyzes the conductance of this hybrid 2D atomic system, as probed by the inherent two-reservoir geometry  [Fig.~\ref{Fig:TransportChannel}]. At this stage, let us highlight a couple of peculiarities introduced by the synthetic ($\lambda$) dimension. First, we emphasize that this synthetic dimension is intimately related to the energy of the system (each ``site'' along $\lambda$ is associated with a harmonic-oscillator level), and hence, it cannot be simply treated as a genuine spatial direction. In particular, there is a built-in chemical-potential bias along the $\lambda$ direction, in the sense that particles privilege the occupation of low-$\lambda$ (i.e.~low-energy) states; this natural bias leads to a subtle interplay with the overall chemical-potential imbalance that is imposed by the two reservoirs to drive current across the channel [Fig.~\ref{Fig:TransportChannel}(a)]. Second, since the system is periodically driven (and thus belongs to the class of Floquet-engineered systems~\cite{Kitagawa2010,GoldmanPRX,Bukov2015,EckardtRMP}), transport properties rely on the underlying quasi-energy spectrum~\cite{Kohler, Kitagawa, Yap}. Altogether, the population of the Floquet eigenstates associated with the inner driven system is non-thermal, but it reflects the thermal population in the undriven reservoirs.

As explained in more detail in the following Sections, these unusual features lead to an effective (fictitious) multi-terminal geometry, which allows us to substantially improve the conductance measurement stemming from the (real) two-reservoir geometry [Fig.~\ref{Fig:TransportChannel}(a)]. As a central result of our work, we demonstrate that a proper state preparation and reservoir configuration can allow for a clear separation of the bulk and edge contributions to  the conductance. In particular, our unusual single-channel setup can be designed so as to fully resolve the quantized Hall conductance associated with chiral edge modes (we recall that this measurement requires at least four terminals in conventional static systems~\cite{Goerbig}). Our work opens new avenues towards the exploration of topological transport in ultracold-atom experiments, through the development of new probing schemes based on synthetic dimensions.

\subsection{Outline}
The paper is organized as follows: The first Section~\ref{Sec:Hall} reviews notions that play an important role in the main part of our study.
We discuss the transport properties of a simple 2D quantum Hall system, the Harper-Hofstadter model~\cite{Hofstadter}, with particular attention to the main differences between the transport measurements that are performed using two-terminal and four-terminal geometries~\cite{Goerbig}.

Section~\ref{Sec:ShakenChannel} introduces the shaken-channel scheme; we discuss the emergence of a synthetic dimension~\cite{Price}, derive an effective time-independent model, and propose a possible implementation using the constriction potential of Fig.~\ref{Fig:TransportChannel}. The transport properties of the effective time-independent model model are studied in two different regimes: in Sec.~\ref{Sec:EffNoAnisotropy}, we make an approximation and map the model to the standard Harper-Hofstadter Hamiltonian; in Sec.~\ref{Sec:EffAnisotropy}, we relax this approximation, and we show that the signatures of quantized transport survive.
We also discuss how the energetic nature of the synthetic dimension naturally leads to an effective multi-terminal geometry, which greatly enriches the measurement based on the (real) two-reservoir geometry.

In Sec.~\ref{Sec:Floquet}, we consider the full time-dependent problem and apply a transport formalism that accurately takes the periodically-driven nature of the system into account~\cite{Kohler, Tsuji, Kitagawa, Yap}. We study the two aforementioned regimes, in Sec.~\ref{Sec:FloquetIsotropy} and in Sec.~\ref{Sec:FloquetFull}, respectively.
Experimental considerations are briefly discussed in Sec.~\ref{Sec:Exp}, and 
conclusions are drawn in Sec.~\ref{Sec:Conclusions}.

\section{Review on quantum Hall transport:~Application to the Harper-Hofstadter model}
\label{Sec:Hall}
Our method builds on the equivalence between energy-resolved two-terminal measurements on a driven one dimensional system, and multi-terminal measurements on a fictitious two-dimensional setup. This Section presents the target model, namely the Harper-Hofstadter model and its topological conductance properties, which will serve as a benchmark for our protocol, as detailed in later Sections.
 
We review the peculiar transport properties associated with the quantum Hall effect, by applying the theoretical framework offered by the Landauer-B\"{u}ttiker formalism, which was originally developed for calculating the conductance in solid-state systems~\cite{Landauer, Buttiker, Imry, Datta}, to the Harper-Hofstadter model~\cite{Hofstadter}. For a review of the Landauer-B\"{u}ttiker formalism and of the recursive Green's function (RGF) method that was used to calculate the conductance shown in our paper, see Appendix~\ref{Sec:TransportMethods}. A generalization of the RGF method~\cite{Kohler, Tsuji, Kitagawa, Yap}, which is specifically tailored to treat time-periodic systems, is presented in Appendix~\ref{Sec:TransportFloquet}.

The main goal of this Section is to distinguish between the transport measurements that result from two-terminal and four-terminal geometries; we will assume zero temperature throughout. We will also discuss how the results of the RGF method can be interpreted in terms of (topological) edge-state transport, based on the Landauer-B\"{u}ttiker formalism~\cite{Gagel95, Gagel96}. The results of this Section will constitute a good basis for understanding the transport properties of the time-independent shaken-channel model [Fig.~\ref{Fig:TransportChannel}], which approximately maps onto the Harper-Hofstadter model (see Section~\ref{Sec:EffNoAnisotropy}).

The Harper-Hofstadter Hamiltonian describes a particle moving in a two-dimensional lattice in the presence of a perpendicular magnetic field~\cite{Hofstadter}
\begin{equation}
\hat{H} =\! -J\sum_{i,j}\Big(e^{\mathrm{i} \varphi j} \ket{i,j}\bra{i+1,j} + \ket{i,j}\bra{i,j+1} + \mathrm{H. c.}\Big).
\label{HHm}
\end{equation} 
This Hamiltonian describes hopping processes taking place between nearest-neighboring sites $(i,j)$ of the lattice, where the indices $i$ and $j$ refer to the two directions ($x$ and $y$), respectively. For fractional values of the flux $\varphi= 2 \pi p/q$, with $q,\,p \in \mathbb{Z}$, the spectrum of the Hamiltonian in Eq.~\eqref{HHm} depicts $q$ bulk bands, which are connected by topologically-protected chiral edge states~\cite{hatsugai1993chern}. These chiral edge states are responsible for the quantized Hall conductance of the system, whenever the Fermi energy lies in a spectral bulk gap~\cite{vonKlitzing, Thouless, Laughlin,hatsugai1993chern}. 
Specifically, in this quantum Hall regime, the longitudinal conductance vanishes, while the transverse (Hall) conductance exhibits robust plateaus~\cite{vonKlitzing}, whose values directly correspond to the number of current-carrying edge states~\cite{macdonald1984edge,hatsugai1993chern}. This behaviour can be understood as the bulk of the system being insulating, while chiral edge currents carry the Hall current. Importantly, the chirality (orientation) of these edge modes around the 2D sample determines the sign of the Hall conductance.

We shall now discuss how these considerations apply to the conductance signal that is extracted from transport measurements using two or four terminals.

\subsubsection{Two-terminal geometry}
\label{Sec:HH2terminal}
\begin{figure}[t]
\centering
\includegraphics[width=1 \columnwidth]{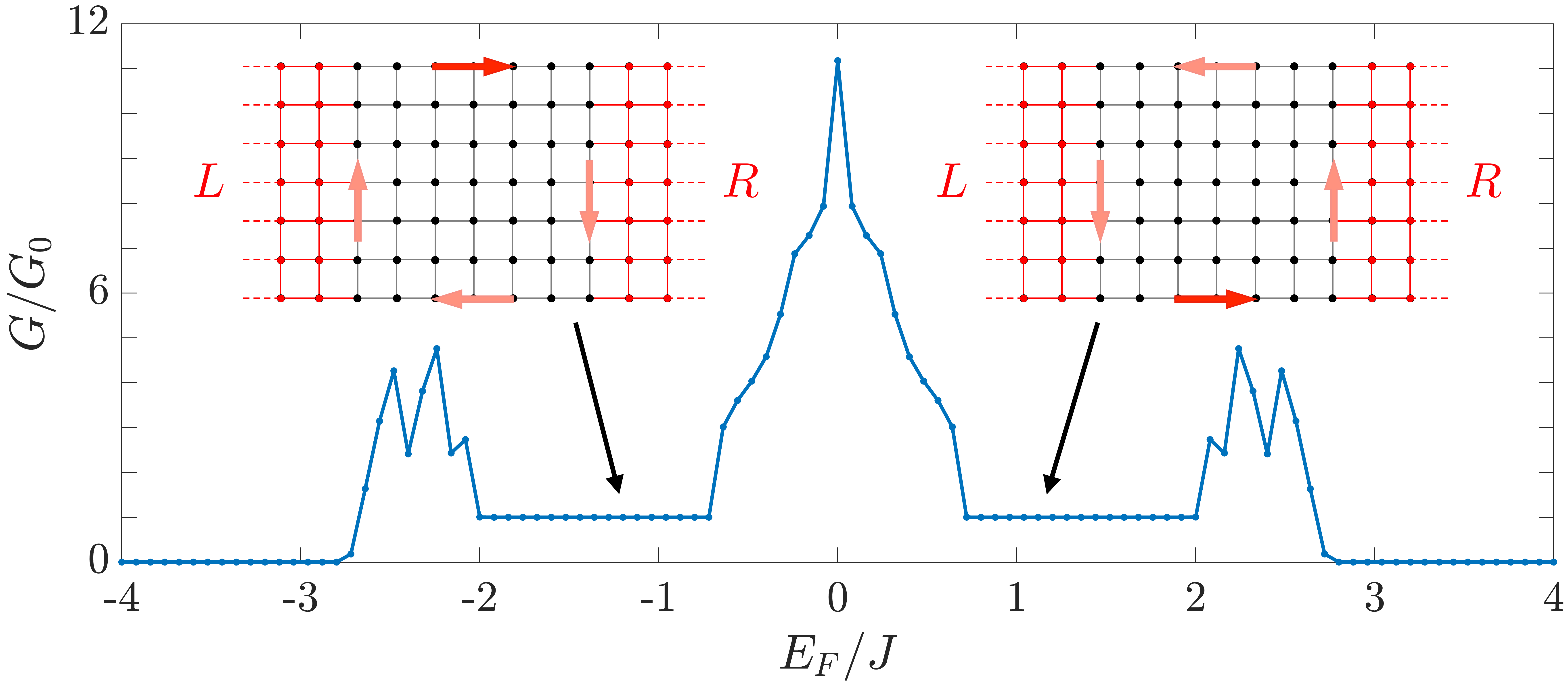}
\caption{(insets) Example of a two-terminal setup with a square lattice geometry of $N_x=N_y=7$ sites along the $x$ and $y$ directions, respectively. The inner system is represented by the black dots, while the red dots identify the sites of the left (L) and right (R) reservoir. The arrows describe the motion of the chiral edge modes present in the model (clockwise at energies $E\!<\!0$ and counter-clockwise at $E\!>\!0$). (main) Conductance $G_{L,R}$ of the Harper-Hofstadter model with magnetic flux $\varphi=2\pi/3$, within the wide-band approximation (see Appendix~\ref{Sec:TransportMethods}), for an input bias applied to the left reservoir; the system size is $N_x\!=\!N_y\!=\!30$. The plateaus are associated with the edge modes illustrated in the insets.}
\label{Fig:HH2terminal}
\end{figure}

We consider a square lattice with $N_x$ and $N_y$ sites along the $x$ and $y$ directions, respectively. This lattice constitutes the inner system to be probed. Then, each site at the left and right end of the inner system is coupled to a reservoir, through a left and right terminal, respectively. An example of such a geometry is shown in Fig.~\ref{Fig:HH2terminal}, where the inner system is represented in black, while the reservoir sites and the terminal links connecting them to the inner system are indicated in red. For simplicity, we have indicated only a few sites of the reservoirs, as the latter are assumed to be very large and to have translational symmetry (this is schematically indicated by dashed lines). The two reservoirs are labelled with $L$ and $R$, respectively, for left and right.
We assume that a small bias is applied to the left reservoir and that the conductance $G_{L,R}$ is measured through a current detected at the right terminal. 
The linear d.c.~current at zero temperature is
$I_{L, R}= G_{L, R}(E_F)  \Big(\mu_L -\mu_R \Big)$, 
where $G_{L, R}(E_F)$ is the conductance of the system at a given Fermi energy $E_F$, and $\mu_{L}$ ($\mu_R$) is the chemical potentials of the $L$ ($R$ ) reservoir~\cite{Imry}.
Following the Landauer formalism~\cite{Landauer}, the conductance is
\begin{align}
G_{L,R}(E_F) \equiv G_0 \sum_m T_{L,R}^{(m)}(E_F),
\label{zeroConductance}
\end{align}
where the sum is over all possible transport channels, and where we considered the case of spinless (single-component) fermions; here $G_0 = 1/h$ denotes the quantum of conductance, and we set $e\!=\!1$ to equally treat charged and neutral particles in this work.
The transmission probability $T_{L,R}^{(m)}(E_F)$ is calculated from the scattering properties of the system following standard procedures, such as the RGF method \cite{Datta, Ryndyk, Lewenkopf, Caroli, ThoulessRGF, Thorgilsson}. More details are given in Appendix~\ref{Sec:TransportMethods}.

The inner system is described by the Harper-Hofstadter model in Eq.~\eqref{HHm}, and we will fix the flux per plaquette to the value $\varphi\!=\!2\pi/3$, for which the spectrum depicts three isolated bulk bands; in this setting, the two spectral gaps host a single chiral edge mode each~\cite{hatsugai1993chern}: the bulk gap at $E\!>\!0$ [resp.~$E\!<\!0$] hosts an edge mode that propagates counter-clockwise [resp.~clockwise] around the 2D lattice. When the Fermi energy $E_F$ lies in the middle of a bulk band, one expects the observation of a metallic behavior: bulk states provide a large set of non-perfectly transmitting channels, which results in a non-quantized conductance across the system. In contrast, when the Fermi energy is set within a band gap, the only channels that are available for transport are provided by the edge modes; in this regime, the conductance is quantized according to the number of edge modes present in the gap (i.e.~one in the present model). Importantly, the chirality of the edge modes (and hence, the sign of the Hall conductance) cannot be identified in a two-terminal geometry~\cite{Goerbig}. Particles populating the ``clockwise'' (resp.~``counter-clockwise'') chiral edge mode flow from the left to the right reservoir by following the top (resp.~bottom) edge; see insets in Fig.~\ref{Fig:HH2terminal}. In both cases, this gives rise to a positive (quantized) conductance between the two reservoirs. In this sense, measuring a quantized conductance $G_{L,R}$ in this two-terminal geometry can only reveal the absolute value of the Hall conductance associated with the underlying 2D lattice system~\cite{Goerbig}. We illustrate this phenomenon in Fig.~\ref{Fig:HH2terminal}, where we plot the conductance $G_{L,R}$ resulting from the RGF method, as a function of the Fermi energy $E_F$ of the inner system (of size $N_x\!=\!N_y\!=\!30$). This plot clearly indicates that the conductance $G_{L,R}$ is quantized and positive whenever the Fermi energy falls within one of the two band gaps of the model [see the plateaus in Fig.~\ref{Fig:HH2terminal}], in agreement with the discussion above. Conversely, this same conductance is found to be not quantized whenever the Fermi energy hits one of the three bulk bands [see the irregular peaks in Fig.~\ref{Fig:HH2terminal}]. 
Summarizing, while this two-terminal measurement cannot capture the chirality of the edge modes (and hence, the sign of the quantized Hall conductance), it does give a clear indication that the system displays perfectly-transmitting channels (i.e.~potential chiral edge modes) within well-defined energy ranges. 

\subsubsection{Four-terminal geometry}
\label{Sec:HH4terminal}

\begin{figure}[t]
\centering
\includegraphics[width=1 \columnwidth]{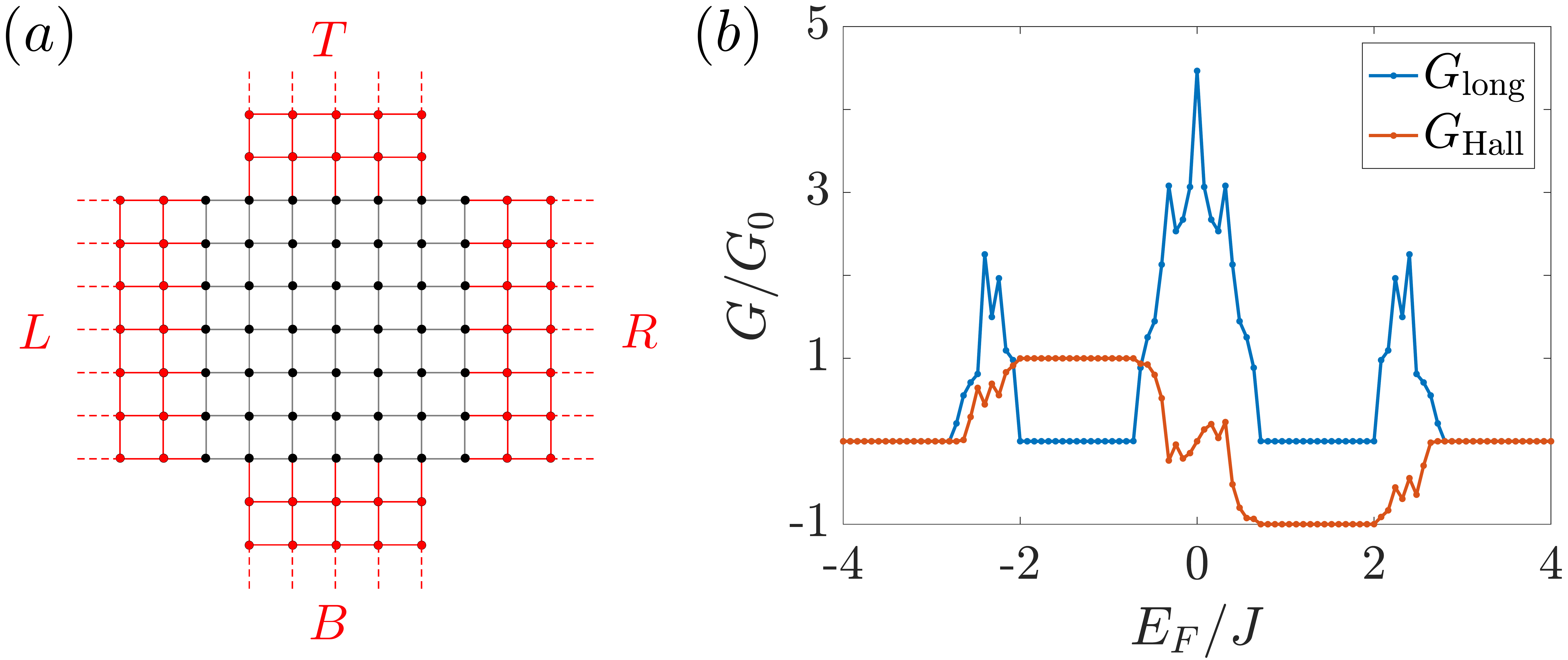}
\caption{(a) Example of a four-terminal setup with a square lattice geometry of $N_x\!=\!N_y\!=\!7$ sites along the $x$ and $y$ directions, respectively. The system is represented by the black dots, while the red dots identify the sites of the left (L), right (R), top (T) and bottom (B) reservoirs. (b) Longitudinal ($G_\text{long} \!=\! G_{L,R}$) and transverse Hall conductance ($G_\text{Hall} \!=\! G_{L, T} - G_{L,B}$) of the Harper-Hofstadter model with magnetic flux $\varphi=2\pi/3$, for an input bias applied to the left reservoir; the system size is $N_x\!=\!N_y\!=\!30$.}
\label{Fig:HH4terminal}
\end{figure}

We now extend the two-terminal geometry by adding two more terminals at the top and at the bottom of the inner system; we label these $T$ and $B$, respectively. As a technical note, we impose that the corner sites of the inner system are only coupled to a single reservoir, which allows one to unambiguously define the terminal regions. An example of such a four-terminal geometry is shown in Fig.~\ref{Fig:HH4terminal}(a). As further illustrated below, this four-terminal configuration allows for a clear and independent identification of the longitudinal and transverse (Hall) conductances of the system; we note that this configuration forms a minimal ``Hall-bar'' setup, as routinely used in solid-state experiments. 

Specifically, the longitudinal conductance $G_\text{long}\!\equiv\!G_{L,R}$ is obtained as in Sec.~\ref{Sec:HH2terminal}, namely, by applying a small bias to the left reservoir and measuring a current at the right terminal.

In order to extract the transverse (Hall) conductance, $G_\text{Hall}$, one needs to analyze the transport that takes place between the top and the bottom reservoirs (while a bias is imposed at the left reservoir, as above). Applying the Landauer-B\"uttiker formalism to the four-terminal configuration~\cite{Buttiker86, Buttiker88}, one finds that the Hall conductance can be obtained as $G_\text{Hall}\!\equiv\!G_{L,T} - G_{L,B}$, which is indeed suitable in the present configuration where the bias is set in the left reservoir. We note that this expression for the Hall conductance is specifically chosen so as to probe the unidirectional transport of chiral edge states, propagating around the 2D quantum Hall system~\cite{Buttiker86}. 

In Fig.~\ref{Fig:HH4terminal}(b), we show the longitudinal conductance $G_\text{long}$ and the Hall conductance $G_\text{Hall}\!\equiv\!G_{L,T} - G_{L,B}$, as obtained using the RGF method described in Appendix~\ref{Sec:TransportMethods}. As before, the magnetic flux is $\varphi\!=\!2\pi/3$ in the inner system and the number of sites is $N_x\!=\!N_y\!=\!30$. In contrast with the result shown in Fig.~\ref{Fig:HH2terminal}, one now obtains a clear signature of the quantum Hall effect~\cite{vonKlitzing, Goerbig}: the longitudinal conductance vanishes and the Hall conductance $G_\text{Hall}$ depicts clear plateaus whenever the Fermi energy $E_F$ falls within a band gap. In particular, the chirality of the propagating edge modes is now clearly identified through the sign of the (Hall) conductance. 

Summarizing, a multi-terminal setup (including at least four terminals) is required to unambiguously measure the quantized Hall conductance of a 2D system, and to fully characterize the nature of the underlying chiral edge modes~\cite{Goerbig}. As will be discussed below [Section~\ref{Sec:EffNoAnisotropy}], the energetic nature of the synthetic dimension that emerges from the shaken-channel system in Fig.~\ref{Fig:TransportChannel} naturally leads to an \emph{effective multi-terminal} configuration, although the constriction-potential only creates two atomic reservoirs: one on each side of the channel. This important observation is at the core of our present proposal.

\section{Transport in a shaken channel: The effective Hamiltonian approach}
\label{Sec:EffectiveTransport}

In this Section, we introduce the mapping between the two-terminal conductance measurements on a driven one-dimensional channel and the more conventional quantum Hall measurement as performed in two-dimensional systems. Specifically, our approach builds on (i) the mapping of a one-dimensional driven channel onto a fictitious two-dimensional lattice system via the concept of synthetic dimension~\cite{Price}; (ii) the interpretation of the two non-driven reservoirs connected to the channel as many fictitious reservoirs connected along the synthetic dimension. The latter provides a natural bias along the synthetic dimension, allowing for a direct measurement of a Hall-like response using a single channel.

\subsection{The model}
\label{Sec:ShakenChannel}

In this Section, we define the shaken-channel model [Fig.~\ref{Fig:TransportChannel}], and describe its transport properties using an effective-Hamiltonian approach. We discuss how a natural synthetic dimension emerges in the problem, and elaborate on how this feature affects the coupling to the reservoirs. In particular, this leads to the notion of ``effective multi-terminal configurations'', which allows for a clear detection of the quantum Hall effect in the shaken-channel model. The effective time-independent Hamiltonian approach will be further validated in the full-time-dependent approach of Section~\ref{Sec:Floquet}.

\subsubsection{The shaken channel}

We consider a non-interacting gas of ultracold fermions (of mass $M$), which are restricted to move within a single channel aligned along the $y$ direction; we focus our attention on the channel, and disregard the reservoirs for now. This system will be described by a single-particle Hamiltonian of the form
\begin{align}
\hat{H}_0=\frac{\hat{p}_x^2}{2M} + \frac{M\omega^2}{2} \hat{x}^2 -  J_y \sum_j \Big[\vert x, j \rangle \langle x, j+1 \vert + \text{H.c.} \Big] .\label{starting_ham}
\end{align}
The first two terms in Eq.~\eqref{starting_ham} describe the motion along the harmonically-confined transverse direction ($x$), with trapping frequency $\omega$, while the last term describes motion along the channel. Here, we have assumed that a deep lattice potential is set along the channel, and that a single-band tight-binding approximation can be made to capture the dynamics along this direction. Then the hopping processes between neighboring orbitals, $\ket{x,j}$ and $\ket{x,j+1}$, are fully characterized by the tunneling parameter $J_y$; here $j=y/a$ refers to the site index along the channel direction, and we set the lattice spacing $a=1$ in the following. 

Inspired by Ref.~\cite{Price}, we subject the tight harmonic confinement to a resonant time-periodic modulation, with frequency $\omega_D\!\approx\!\omega$, 
\begin{equation}
\hat{V}(t) = \kappa \hat{x} \cos\left(\omega_D t + \varphi j \right) ,
\label{timemod}
\end{equation}
which corresponds to shaking the channel along the transverse ($x$) direction. Importantly, the modulation in Eq.~\eqref{timemod} includes a phase $\theta(y)\!=\!\varphi j$, which explicitly depends on the channel direction ($y$). In the following, we assume that such a modulation is only active on a (substantial) part of the channel, so that there are intermediate regions that adiabatically connect the non-shaken reservoirs to the shaken channel (i.e.~the inner system).

Following Ref.~\cite{Price}, we write the total time-dependent Hamiltonian $\hat{H}(t)\!=\!\hat{H}_0+\hat{V}(t)$ in the basis formed by the harmonic oscillators states, $\ket{\lambda,j}$, where $\lambda$ refers to the discrete harmonic levels associated with the transverse ($x$) direction:
\begin{align}
\hat{H}(t)&= \sum_{\lambda,j}\Big[ \omega \lambda \ket{\lambda,j}\bra{\lambda,j} - \Big(J_y  \ket{\lambda,j}\bra{\lambda,j+1}+ \text{H.c.}\Big)\Big]\notag\\
& + \sum_{\lambda,j} 2 \cos\left(\omega_D t + \varphi j\right) \times \label{Hlambda} \\
& \qquad \Big( J_{\lambda} \ket{\lambda,j}\bra{\lambda+1,j} + J_{\lambda+1}  \ket{\lambda+1,j}\bra{\lambda,j}  \Big),
\notag
\end{align}
where $J_\lambda=\kappa\sqrt{\lambda/8M\omega}$. In the high-frequency regime ($\omega_D\!\approx\!\omega\!\rightarrow\infty$), an effective model can be obtained in the frame rotating at the shaking frequency, by invoking the rotating-wave approximation ~\cite{GoldmanPRX, EckardtRMP, Price}:
\begin{align}
\hat{H}_\text{eff}& = \sum_{\lambda,j}\Big[ \Delta \lambda  \ket{\lambda,j}\bra{\lambda,j} - \Big(J_y  \ket{\lambda,j}\bra{\lambda,j+1} + \text{H.c.} \Big) \Big] \notag\\
& - \sum_{\lambda,j} \Big( J_\lambda e^{\mathrm{i}\varphi j} \ket{\lambda,j}\bra{\lambda+1,j} + \text{H.c.} \Big) , 
\label{Heff}
\end{align}
where $\Delta\!=\! \omega_D -\omega$ is the detuning between the trap and drive frequencies; in the following, we shall consider a resonant drive with $\Delta\!=\!0$. As previously discussed in Ref.~\cite{Price}, the rotating-wave approximation and, hence, the description based on the effective Hamiltonian in Eq.~\eqref{Heff}, breaks down whenever $|J_\lambda| \gtrsim \omega/4$. This limits the number of states $\lambda$ that are available along the synthetic dimension, for a given ratio $\kappa l_H/\omega$, where $l_H\!=\!1/\sqrt{M\omega}$ is the harmonic length (which sets a natural length scale in the problem). Under these approximations, the effective time-independent Hamiltonian in Eq.~\eqref{Heff} corresponds to a 2D tight-binding model, defined on a square lattice in the $\lambda-y$ plane, with inhomogeneous and anisotropic hopping strengths $(J_{\lambda}, J_y)$. Furthermore, this 2D-lattice model includes a uniform artificial magnetic field, which corresponds to having $\varphi$ quanta of flux per plaquette. Summarizing, the model realizes the Harper-Hofstadter model in Eq.~\eqref{HHm}, with inhomogeneous and anisotropic hopping parameters. We note that the optical lattice set along the channel direction is not a crucial ingredient, as removing it would result in a model of ``quantum-Hall wires'' with similar properties~\cite{Kane, Budich}. 

\subsubsection{Connecting the shaken channel to reservoirs}

Our proposal builds on the observation that the shaken channel described above is naturally connected to two reservoirs, as illustrated in Fig.~\ref{Fig:TransportChannel}. Specifically, the total system is constituted of an inner 2D system realizing the anisotropic Harper-Hofstadter model (defined in the fictitious $\lambda-y$ plane), which is connected to two non-shaken reservoirs. The latter are attached at both ends of the channel, namely, they connect to the 2D inner system at $y\!=\!0$ and $y\!=\!L_y$. 

In the following, we will consider that a chemical potential imbalance can be applied between the two reservoirs, so as to force the atoms to move from the left side of the channel to the right side [Sec.~\ref{Sec:Hall}]; the relative particle difference then provides a measurement of the system's conductance~\cite{Krinner_Rev}. Since the effective model describing the 2D inner system corresponds to the Harper-Hofstadter-like model in Eq.~\eqref{Heff}, one expects that the resulting transport properties should be reminiscent of those presented in Sec.~\ref{Sec:HH2terminal} (which discussed the conductance of the Harper-Hofstadter model as probed by a two-terminal geometry). 

This naive prediction is based on the assumption that the synthetic ($\lambda$) direction can be treated as a genuine spatial direction:~specifically, it assumes that the reservoirs inject particles in the inner system, at $y\!=\!0$, in a $\lambda$-independent manner. 
However, the time-modulated (shaken) channel maps onto the model in Eq.~\eqref{Heff} in a frame rotating at the shaking frequency, while the reservoirs have a thermal distribution in the laboratory frame. The transport measurement therefore needs to account for the fact that, in the laboratory frame, the $\lambda$ direction is the energy axis and the channel is explicitly time dependent. As we show below, this actually provides a novel route for the investigation of the topological features of the HH model. 

\subsection{Simple effective model: Homogeneous hopping along the synthetic dimension}
\label{Sec:EffNoAnisotropy}
For the sake of clarity, let us first simplify the analysis of the time-independent effective model in Eq.~\eqref{Heff}, by  neglecting the inhomogeneity in the hopping along the synthetic dimension $\lambda$; specifically, we substitute $J_{\lambda}\!\rightarrow\!J_1\!\equiv\!\kappa \sqrt{1/8M\omega}$. The corresponding effective model that describes the inner system then reads
\begin{align}
\hat{H}_\text{eff}^{\text{HH}} = &-J_y\sum_{\lambda,j} \Big( \ket{\lambda,j}\bra{\lambda,j+1} + \text{H.c.}\Big) \notag\\
&-J_1\sum_{\lambda,j} \Big(  e^{\mathrm{i}\varphi j}  \ket{\lambda,j}\bra{\lambda+1,j} + \text{H.c.}\Big),
\label{Heffisotropic}
\end{align}
and it exactly maps onto the Harper-Hofstadter model when $J_y=J_1$. We note that the approximation of constant $J_\lambda$ is only valid in the limit of large $\lambda\!\gg\!1$, which is a priori problematic since we anticipate that low-$\lambda$ states should substantially contribute to transport (we remind the reader that transport is dominated by chiral edge modes in the quantum-Hall regime, and that $\lambda\!=\!0$ defines a clear edge along the synthetic dimension). In this sense, the results presented in this Section only aim to provide a general intuition on the transport that takes place in this synthetic 2D system. The analysis of the full (inhomogeneous) effective Hamiltonian will be postponed to Sec.~\ref{Sec:EffAnisotropy}, while the properties of the full time-dependent model (including the application of the Floquet-Landauer approach to transport) will be presented in Sec.~\ref{Sec:Floquet}.

Although the synthetic direction $\lambda$ introduced above is semi-infinite, we note that, in practice, the anharmonicity of the confinement realized in ultracold-atom experiments can produce a natural (soft) boundary. Besides, a box-type potential could be applied on top of the harmonic trap to further confine the particles along the transverse direction, which would result in a sharp edge in the $\lambda$ direction (at a desired $\lambda_\text{cut}$). In the following, we introduce a cutoff along the synthetic direction $\lambda_\mathrm{cut}\!\gg\!1$, which is convenient for performing numerical simulations on a finite system.

\subsubsection{Two-terminal geometry: Adding details to the reservoirs} 
\label{Sec:Eff2Terminal}

\begin{figure}[t]
\centering
\includegraphics[width=1 \columnwidth]{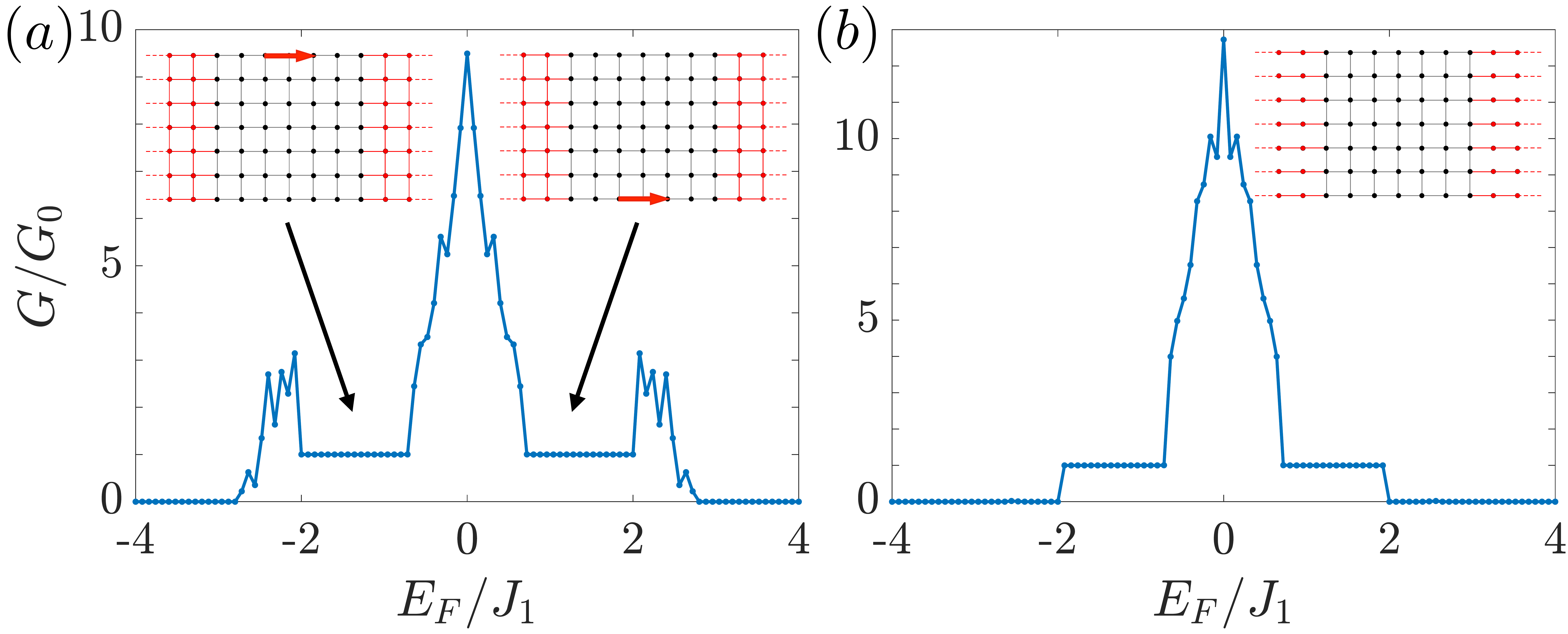}
\caption{Two-terminal conductance of the isotropic effective model in Eq.~\eqref{Heffisotropic}. Parameters are $\kappa=0.01\omega/l_H$, $J_y=J_1$, $\varphi=2\pi/3$, $\lambda_\mathrm{cut}=29$, $N_y=30$. The sites in the reservoirs are always horizontally coupled $J_y^r=J_y$, while vertical couplings in the reservoirs are (a) $J_1^r\!=\!J_1$, such that the wide-band approximation is valid, or (b) $J_1^r\!=\!0$, for which the wide-band approximation breaks down.}
\label{Fig:RealisticCoupling}
\end{figure}

As a first step, we treat the synthetic dimension as a genuine spatial dimension. Under this assumption, the conductance of the model in Eq.~\eqref{Heffisotropic} can be calculated as in Section~\ref{Sec:HH2terminal}, namely, by treating the full system (inner part and reservoirs) as a standard two-terminal problem. A sketch of this simple configuration is shown in the inset of Fig.~\ref{Fig:RealisticCoupling}(a).

We now check whether the details of the reservoirs influence the calculation of the conductance. This is different from applying the wide-band approximation described in Appendix ~\ref{Sec:TransportMethods}, as was previously considered for the calculation shown in Fig.~\ref{Fig:HH2terminal}. As a first assumption, we take the Hamiltonian describing the reservoir as in Eq.~\eqref{Heffisotropic} but with $\varphi \!=\!0$, i.e.~the effective magnetic field is assumed to be absent in the reservoirs. Furthermore, we will first (naively) assume that the hopping amplitudes are uniform and isotropic throughout the entire system: $J_y^r\!=\!J_y$ and $J_1^r\!=\!J_1$, where the superscript $r$ refers to the reservoirs. The resulting conductance plot is shown in Fig.~\ref{Fig:RealisticCoupling}(a), which naturally shares the same features as those previously displayed in Fig.~\ref{Fig:HH2terminal} [Section~\ref{Sec:HH2terminal}]. In particular, the plateau at $E_F\!>\!0$ in Fig.~\ref{Fig:RealisticCoupling}(a) is attributed to the counter-clockwise propagating edge mode, which is associated with the spectrum of the Harper-Hofstadter Hamiltonian in Eq.~\eqref{Heffisotropic}, and which is localized at $\lambda\!=\!0$. Conversely, the plateau at $E_F\!<\!0$ is associated with the clockwise propagating edge mode localized at $\lambda\!=\!\lambda_\text{cut}$; see inset of Fig.~\ref{Fig:RealisticCoupling}(a). 

We stress that the description used above for the reservoirs ($J_y^r\!=\!J_y$ and $J_1^r\!=\!J_1$) is not compatible with the actual scheme described in Section~\ref{Sec:ShakenChannel}. Indeed, since the temporal modulation only acts on the inner part of the system (or more precisely, on a substantial part of the transport channel), a more accurate description consists in setting $J_1^r\!=\!0$ in the junction regions connecting the inner system to the reservoirs (noting that the harmonic-oscillator states $\lambda$ are indeed decoupled in the absence of the time-modulation); see the sketch in the inset of Fig.~\ref{Fig:RealisticCoupling}(b). As shown in the conductance plot of Fig.~\ref{Fig:RealisticCoupling}(b), the quantized plateaus remain unaffected by this modification of the reservoirs properties [compare Figs.~\ref{Fig:RealisticCoupling}(a) and (b)]. 

As a technical remark, we note that the bulk-band response at $|E_F/J_1|>2$ is dramatically suppressed in Fig.~\ref{Fig:RealisticCoupling}(b), which is due to the breakdown of the wide-band approximation: when setting $J_1^r\!=\!0$, the bandwidth associated with the reservoirs is of the order of $W^r\!\approx\!2 J_y^r$, which is smaller than the bandwidth of the inner system, $W\!=\!4J_1\!=\!4 J_y^r$, in the situation considered in Fig.~\ref{Fig:RealisticCoupling}(b). We have checked that the bulk response of Fig.~\ref{Fig:HH2terminal} is indeed recovered when setting $J_y^r\!\gtrsim\!J_y\!=\!J_1$, i.e.~when reaching a regime where the wide-band approximation is again satisfied. Importantly, we verified that the details of the reservoirs do not break the robustness of the quantized plateaus, as soon as the wide-band approximation is fulfilled. In the remainder of the paper, we will always treat the reservoirs in the wide-band approximation.

\subsubsection{Effective multi-terminal geometry} 
\label{Sec:EffMultiTerminal}

\begin{figure}[t]
\centering
\includegraphics[width=1 \columnwidth]{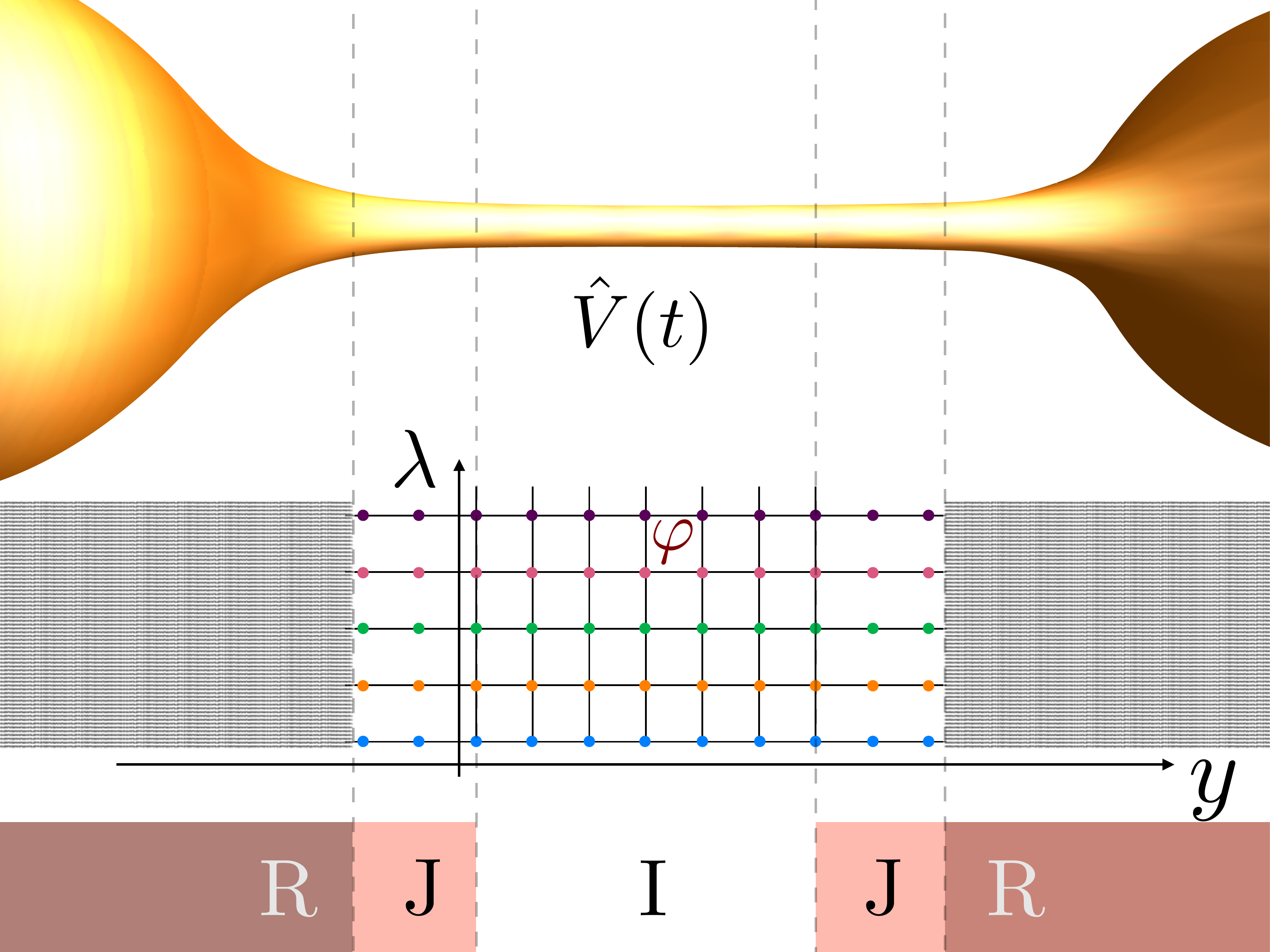}
\caption{Connecting the fictitious 2D system to two reservoirs: (I) the inner system where the time-modulation $\hat V (t)$ is applied (resulting in a fictitious 2D lattice defined in the $\lambda\!-\!y$ plane), (R) the two reservoirs, (J) the two junction regions (where coupling between the $\lambda$ states is absent). Particles leave a reservoir to enter a non-shaken junction region (where the populations of the $\lambda$ states only rely on the Fermi-Dirac distribution associated with the reservoir); particles then leave the junction to enter the inner system, in an adiabatic manner. }
\label{Fig:Junction}
\end{figure}

As a crucial step in the description and understanding of our scheme, we now take the energetic nature of the synthetic dimension into account. To do so, we analyze how particles are injected from the left reservoir into the inner shaken channel by partitioning the whole system into three connected parts: (a) the inner system (2D lattice in the fictitious $\lambda-y$ plane), (b) the two reservoirs, (c) the two junction regions that connect the inner system to the reservoirs; see Fig.~\ref{Fig:Junction}. It is reasonable to assume that the junction regions can be treated as discrete harmonic levels of energy $\hbar \omega$, which are populated following a thermal (Fermi-Dirac) distribution set by the reservoirs: particles in a given $\lambda$ state are injected if $\lambda \hbar \omega\!<\!E_F$, and holes are injected for $\lambda \hbar \omega\!>\!E_F$. In this sense, the reservoir injects more particles in the low-$\lambda$ states (bottom of the hybrid 2D inner system) than in the high-$\lambda$ states:~the reservoir is not ``connected'' uniformly along the synthetic dimension, and there is an effective chemical-potential bias along the $\lambda$ direction. A simple way to include this unusual feature in our effective-Hamiltonian description consists of splitting up the left and right reservoirs into many (fictitious) reservoirs, all aligned along the synthetic $\lambda$ direction with varying chemical potentials (depending on their location along the synthetic dimension); it is the aim of the two following paragraphs to study the resulting transport properties. We point out that this effective bias along the $\lambda$ direction can be naturally controlled experimentally via the overall chemical potential and temperature of the reservoirs.

As before, we still assume that the chemical potentials are set such that transport is driven from the left part to the right part of the system. Besides, in the following paragraphs, we will assume that hopping is allowed in the reservoirs along the synthetic dimension ($J_1^r\!=\!J_1$); this choice can be modified in order to reach an even finer description [i.e.~by setting $J_1^r\!=0$; see discussion of Section~\ref{Sec:Eff2Terminal}]. Other transport configurations will also be briefly discussed below.

\begin{figure}[t]
\centering
\includegraphics[width=1 \columnwidth]{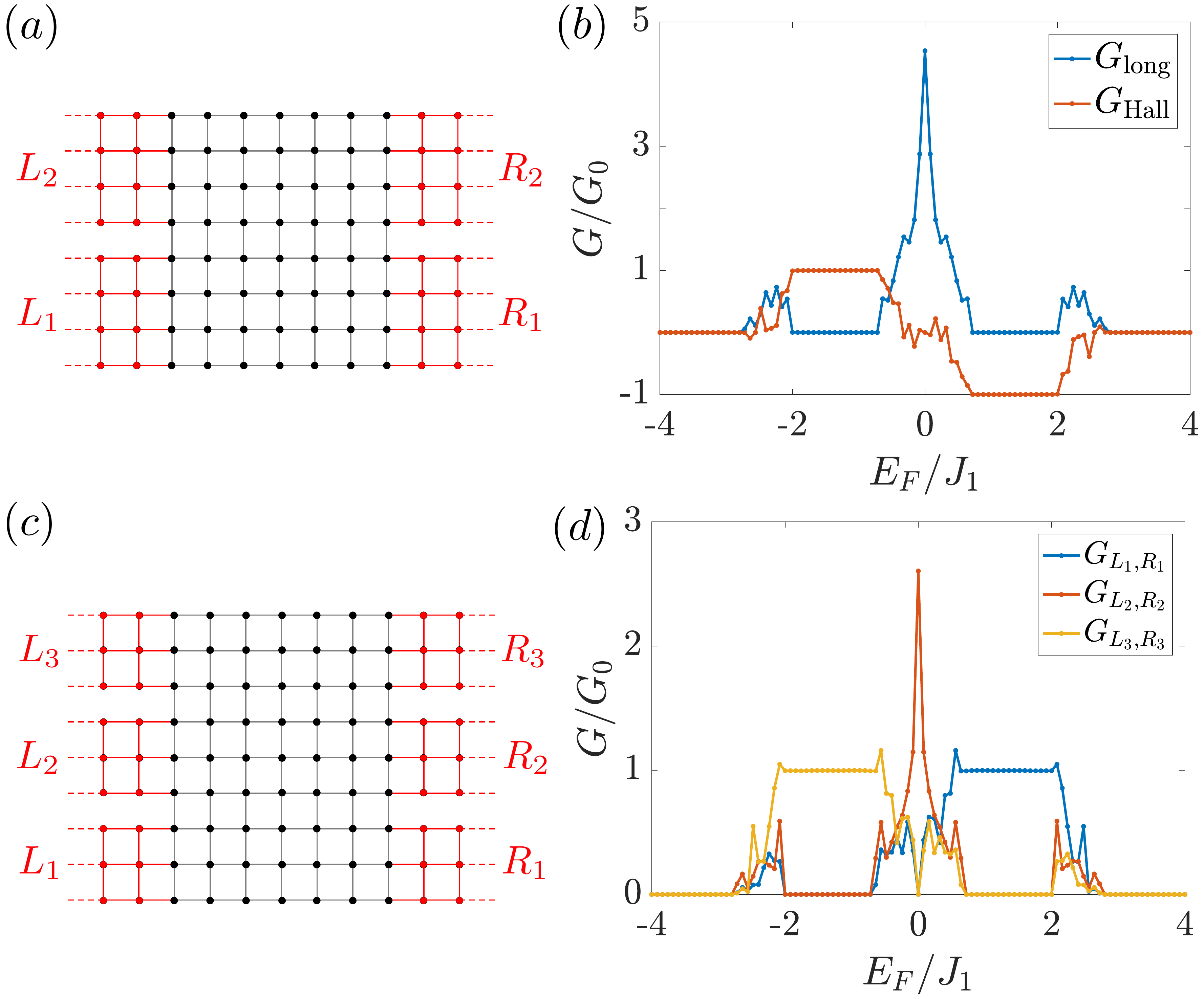}
\caption{(a) Effective four-terminal geometry and (b) the resulting longitudinal and Hall conductances. (c) Effective six-terminal geometry and (d) the resulting conductances $G_{L_i, R_i}$ for $i\!=\!1,2,3$.
Parameters are $\kappa\!=\!0.01\omega/l_H$, $J_y\!=\!J_1$, $\varphi\!=\!2\pi/3$, $\lambda_\mathrm{cut}\!=\!29$, $N_y=30$. }
\label{Fig:EffMultiterminal}
\end{figure}

\paragraph{Effective four-terminal geometry}
As a first approximation, we consider that the initial 2-reservoir configuration can be split into an \emph{effective four-terminal} geometry:~the inner system is coupled to two effective reservoirs on the left (labelled by $L_{1,2}$) and by two effective reservoirs ($R_{1,2}$) on the right; see the sketch in Fig.~\ref{Fig:EffMultiterminal}(a). We point out that this setting corresponds to a rearrangement of the more standard four-terminal geometry previously discussed in Sec.~\ref{Sec:HH4terminal} [Fig.~\ref{Fig:HH4terminal}(a)]. Importantly, it turns out that this unusual (effective) four-terminal geometry allows for a clear measure of the Hall and longitudinal conductances, as we now explain. 

Figure~\ref{Fig:EffMultiterminal}(b) shows the conductances, as obtained from the RGF method for two different configurations: the longitudinal conductance was calculated as $G_\text{long}\!\equiv\!G_{L_1,R_2}$, while the transverse (Hall) conductance was evaluated as $G_\text{Hall}\!\equiv\!G_{L_2,R_2}\!-\!G_{L_1,R_1}$. These choices can be explained based on simple arguments. Firstly, as previously discussed in Sec.~\ref{Sec:HH4terminal}, the longitudinal conductance results from the contribution of the many extended bulk states, and since the transport taking place between the terminals $L_1$ and $R_2$ in Fig.~\ref{Fig:EffMultiterminal}(a) necessarily involves bulk states, it is thus legitimate to define the longitudinal conductance as $G_{L_1,R_2}$ in this context. Secondly, in quantum-Hall systems, the Hall conductance can be attributed to the contribution of the edge modes. We note that the transport between the terminals $L_1$ and $R_1$ in Fig.~\ref{Fig:EffMultiterminal}(a) can be attributed to the propagation of a counter-clockwise edge mode along the bottom edge as well as to bulk states and similarly that the transport between the terminals $L_2$ and $R_2$ involves the bulk and a clockwise edge state following the top edge. Consequently, the difference $G_{L_2,R_2}\!-\!G_{L_1,R_1}$ allows one to reveal the edge-current contribution, and thus, the Hall conductance; a more rigorous derivation can be obtained based on the Landauer-B\"uttiker formalism~\cite{Buttiker86, Buttiker88}.

\paragraph{Effective six-terminal geometry}
One can further refine the model by considering an effective \emph{six-terminal configuration}, where the main (physical) reservoirs are now split into three effective reservoirs each; we will denote these $L_{1,2,3}$ on the left and $R_{1,2,3}$ on the right, respectively; see the sketch in Fig.~\ref{Fig:EffMultiterminal}(c). As compared to the four-terminal configuration [Fig.~\ref{Fig:EffMultiterminal}(a)], the six-terminal geometry allows for an even more direct detection of the bulk and edge contributions to transport. Indeed, $G_{L_2, R_2}$ reflects the transport taking place in the bulk, and hence provides an accurate probe of the longitudinal conductance, whereas $G_{L_1, R_1}$ [resp.~$G_{L_3, R_3}$] reflects the transport associated with the counter-clockwise [resp.~clockwise] edge mode propagating along the bottom [resp.~top] edge. This is demonstrated in Fig.~\ref{Fig:EffMultiterminal}(d), which shows the corresponding conductances, and which indeed reproduces the expected features of the Hall and longitudinal conductances. As a technical remark, we note that $G_{L_1, R_1}$ and $G_{L_3, R_3}$ also show a weak contribution of the bulk states, in the vicinity of the band edges.

This construction of an effective multi-terminal geometry can be straightforwardly extended to the limit where each row of ``sites'' at a given $\lambda$ is connected to a terminal $L_\lambda$ (resp. $R_\lambda$) on the left (resp. right), see Fig.~\ref{Fig:TwoTerminalsAnistropy}(c).
In this extreme case, the conductances $G_{L_{\lambda=0}, R_{\lambda=0}}$ and $G_{L_{\lambda=\lambda_\text{cut}}, R_{\lambda=\lambda_\text{cut}}}$ would isolate the chiral-edge-mode contributions, while the others $G_{L_\lambda, R_\lambda}$ would capture the contributions from the bulk states only. 

\paragraph{Other configurations}
We point out that other transport configurations can be envisaged. For instance, the motion taking place along the synthetic dimension could be probed by analyzing the conductance associated with two fictitious terminals located on a given side of the channel (e.g. $G_{L_\lambda,L_{\lambda'}}$ or $G_{R_\lambda,R_{\lambda'}}$).
Such a motion along $\lambda$ would physically correspond to a heat transport~\cite{Moskalets} within a given (real) reservoir. Besides, the detuning $\Delta$ defined in Eq.~\eqref{Heff} could be used to generate an artificial electric field aligned along the synthetic dimension, hence offering an additional control parameter to the transport experiment.

\subsection{Complete effective model with $\lambda$-dependent hopping} 
\label{Sec:EffAnisotropy} 
So far, we have described the shaken-channel model in terms of the simplified effective Hamiltonian in Eq.~\eqref{Heffisotropic}, namely, the Harper-Hofstadter model [Sec.~\ref{Sec:Hall}] with isotropic and homogeneous hopping ($J_y\!=\!J_1$). This allowed us to analyze how conductance measurements are modified as one changes the reservoirs configuration, offering a first important step in our understanding of how transport takes place in the presence of a synthetic dimension [Section~\ref{Sec:EffNoAnisotropy}].

We now go beyond these studies, and consider the complete effective model in Eq.~\eqref{Heff}, by taking the inhomogeneous hopping along the synthetic dimension ($J_{\lambda}$) into account; as before, we take the resonant-drive limit and set $\Delta\!=\!0$.

\begin{figure}[t]
\centering
\includegraphics[width=1 \columnwidth]{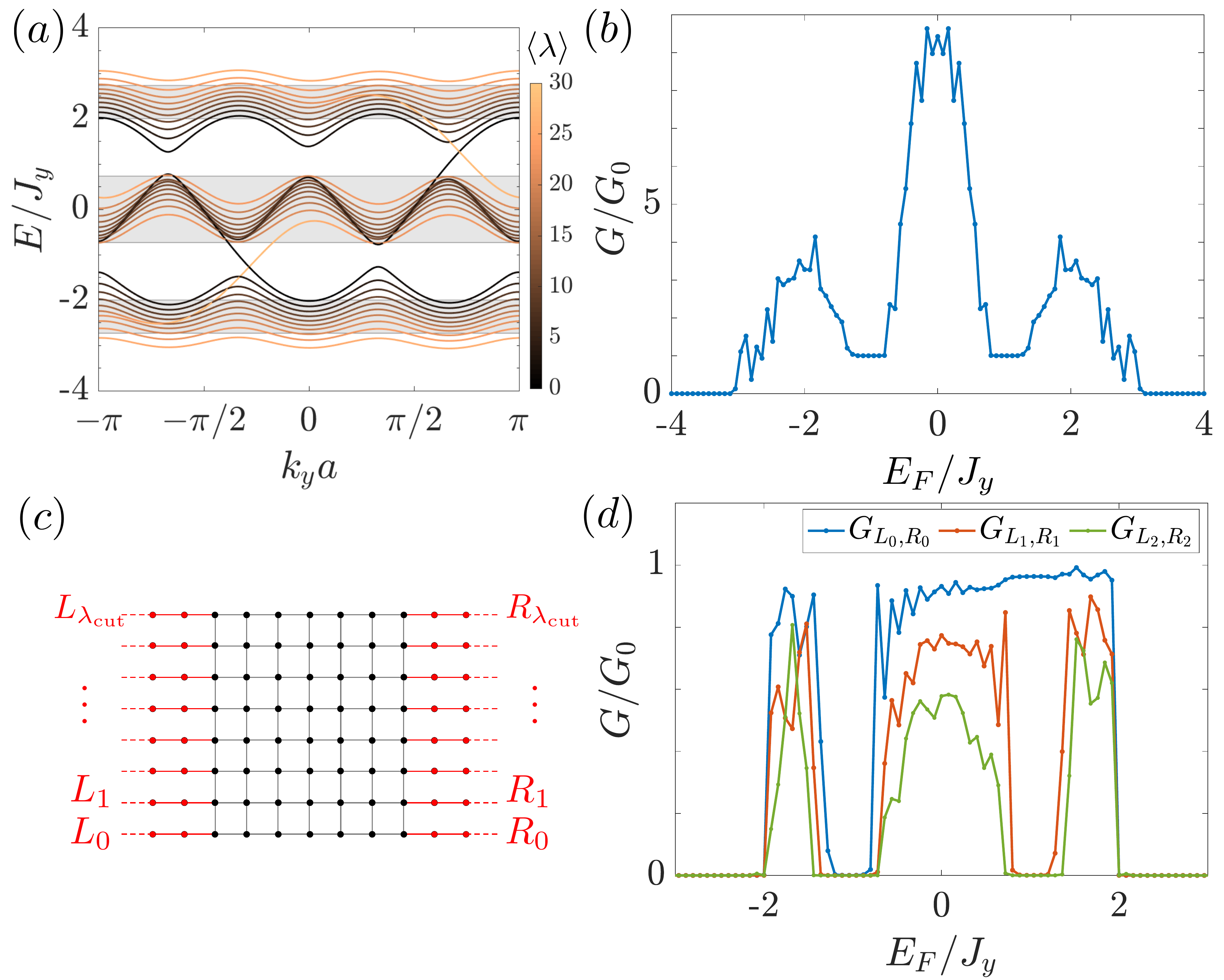}
\caption{(a) Energy dispersion of the complete effective model with $\lambda$-dependent hopping [Eq.~\eqref{Heff}], for periodic boundary conditions along $y$, and $\lambda_\mathrm{cut}\!=\!30$. The color scale, indicating the mean position $\langle \lambda \rangle$ of the eigenstates along the synthetic dimension, is chosen so as to highlight low-$\lambda$ states (darker colors). The shaded regions correspond to the bulk bands of the homogeneous and isotropic model of Sec.~\ref{Sec:EffNoAnisotropy}. (b) Two-terminal conductance of the complete effective model in Eq.~\eqref{Heff}.
(c) Sketch of the effective multi-terminal geometry with $N\!=\!0, \dots \lambda_\text{cut}$, and (d) the resulting conductances $G_{L_\lambda,R_\lambda}$ for $\lambda\!=\!0,1,2$.
Parameters are $\kappa=0.01\omega/l_H$, $J_y=4 J_1$, $\varphi=2\pi/3$, $\lambda_\mathrm{cut}=30$.}
\label{Fig:TwoTerminalsAnistropy}
\end{figure}

As a first step, we calculate the energy spectrum for this effective model, in view of identifying the energy ranges that correspond to the edge modes (and bulk gaps); these ranges will then correspond to the quantized plateaus depicted by the Hall conductance when plotted as a function of the Fermi energy. Following Ref.~\cite{Price}, we diagonalize the Hamiltonian in Eq.~\eqref{Heff} in a gauge where translational symmetry is recovered along the $y$ direction. Applying periodic boundary conditions along the $y$ direction and setting $\lambda_\mathrm{cut}\!=\!30$, one obtains the spectrum shown in Fig.~\ref{Fig:TwoTerminalsAnistropy}(a); the color scale indicates the mean position $\langle\lambda\rangle$ of the eigenstates along the synthetic dimension. The main effect of the inhomogeneous and anisotropic hopping is to increase the bandwidth of the spectrum, and to reduce the size of the bulk gap that hosts the topological edge modes. 
The band width of the complete effective model can be compared with the one of the homogeneous and isotropic model, indicated as gray regions in Fig.~\ref{Fig:TwoTerminalsAnistropy}(a).
For our choice of parameters $J_y = 4 J_1$, these modifications are not too severe in the sense that chiral edge modes are still present in reasonably large bulk gaps (of order $J_y$); we also note that the group velocity (and chirality) of the edge modes are still preserved. For example, the edge mode located in the upper gap, and which propagates from left to right (positive group velocity along $y$), is localized at $\lambda \!=\!0$, while in the lower gap, this mode is localized at $\lambda \!=\! \lambda_\text{cut}$.

We now calculate the conductance of this effective-Hamiltonian system, based on a simple two-terminal configuration, which allows for a direct comparison with the results previously presented in Fig.~\ref{Fig:RealisticCoupling}(a); hence, for clarity, we first neglect the energetic nature of the synthetic dimension in this part of the study. The results are presented in Fig.~\ref{Fig:TwoTerminalsAnistropy}(b), which shows the conductance calculated using the RGF method. Importantly, the plateaus associated with the chiral edge modes (at $\lambda\!=\!0$ and $\lambda\!=\!\lambda_\text{cut}$) are still visible in this more realistic (anisotropic) model. We also notice that the size of the plateaus, which is indicative of the band-gaps displayed in Fig.~\ref{Fig:TwoTerminalsAnistropy}(a), is reduced compared to the isotropic case shown in Fig.~\ref{Fig:RealisticCoupling}(a).

Finally, one can include the energetic nature of the synthetic dimension into the description, by following the effective-multi-terminal construction of Section~\ref{Sec:EffMultiTerminal}, shown in Fig.~\ref{Fig:TwoTerminalsAnistropy}(c). The conductance $G_{L_\lambda ,R_\lambda}$ for the lowest three terminals $\lambda\!=\!0, 1, 2$ is plotted in Fig.~\ref{Fig:TwoTerminalsAnistropy}(d), which  indicates that the edge modes of the realistic model can be unambiguously identified through the multi-terminal geometry of Section~\ref{Sec:EffMultiTerminal}.

\section{Full time-dependent problem: A Floquet-Landauer approach to transport}  
\label{Sec:Floquet}
In the previous Sections, we have analyzed the conductance of the shaken-channel model using an effective-Hamiltonian approach; there, traditional tools of quantum transport were directly applied to time-independent Hamiltonians, which allowed us to explore the peculiarities introduced by the synthetic dimension. In particular, we discussed how the emergent notion of an ``effective multi-terminal configuration'' allows for a clear detection of the transverse (Hall) and longitudinal conductances in a single atomic wire, hence revealing the quantum Hall effect in a minimal cold-atom setting.

In this Section, we now build on these results to analyze the full time-dependent problem, using the Floquet-Landauer-approach to transport described in Appendix~\ref{Sec:TransportFloquet}. The aim of this Section is to fully validate the main result of this article, namely, that the quantized Hall conductance associated with chiral edge modes can be extracted from the shaken-channel model displayed in Fig.~\ref{Fig:TransportChannel}.

Consider the Schr\"{o}dinger equation associated with the full time-dependent Hamiltonian in Eq.~\eqref{Hlambda}, which was expressed in the basis of the harmonic-oscillator-states:
\begin{align}
\mathrm{i} \partial_t \psi_{\lambda,j}(t) =& \sum_{\lambda,j}\Big[  \omega \lambda \psi_{\lambda,j}(t) -J_y \Big( \psi_{\lambda,j+1}(t) + \psi_{\lambda,j-1}(t)\Big)\Big] \notag \\
&+ \sum_{\lambda,j} 2 J_1 \cos\left(\omega t + \varphi j \right)   \label{Schroedinger} \\
& \qquad \times \Big( \sqrt{\lambda} \, \psi_{\lambda-1,j}(t) +\sqrt{\lambda+1}\, \psi_{\lambda+1,j}(t) \Big),\notag
\end{align}
where $\psi_{\lambda,j}(t)$ is the wavefunction of a particle in the state $\ket{\lambda,j}$; as previously, we explicitly set the drive frequency on resonance $\omega_D\!=\!\omega$. The time-periodic system in Eq.~\eqref{Schroedinger} is associated with a quasienergy (Floquet) spectrum, which can be defined in a restricted range $\varepsilon\!\in\![-\omega/2, \omega/2]$; this spectrum can also be represented in an extended-zone scheme, $E = \varepsilon + n \omega$, where the integer $n$ refers to the repeated multiplicities. As discussed in Ref.~\cite{Eckardt2015}, it is convenient to treat such periodically-driven systems in an extended (Floquet) Hilbert space, which explicitly takes these multiplicities into account. We then expand the wavefunction into its Fourier components, 
\begin{equation}
\psi_{\lambda,j}(t) = \sum_{n=-n_F}^{n_F} \psi^{(n)}_{\lambda,j} e^{\mathrm{i}n \omega  t},
\label{Fourierexpand}
\end{equation}
where we have truncated the series up to $2 n_F+1$ modes, and where $\psi^{(n)}_{\lambda,j}$ denotes the time-independent Fourier amplitudes; the number of modes $n_F$ can be chosen so as to reach convergence of the numerical observables. Substituting Eq.~\eqref{Fourierexpand} into Eq.~\eqref{Schroedinger}, and isolating the components proportional to $ e^{\mathrm{i}n \omega  t}$, we obtain the Fourier component $\mathbf{H}_n$ of the Hamiltonian in Eq.~\eqref{Hlambda} (see Appendix~\ref{Sec:TransportFloquet}):
\begin{align}
&\omega \left(\lambda + n\right) \psi^{(n)}_{\lambda,j} - J_y \Big(\psi^{(n)}_{\lambda,j+1} + \psi^{(n)}_{\lambda,j-1} \Big) \label{FourierElement} \\
&+J_1 e^{\mathrm{i} \varphi j}  \Big( \sqrt{\lambda} \psi^{(n-1)}_{\lambda-1,j} + \sqrt{\lambda+1} \psi^{(n-1)}_{\lambda+1,j} \Big) \notag \\
&+J_1 e^{-\mathrm{i} \varphi j}  \Big( \sqrt{\lambda} \psi^{(n+1)}_{\lambda-1,j} + \sqrt{\lambda+1} \psi^{(n+1)}_{\lambda+1,j} \Big) \equiv \mathbf{H}_n \psi^{(n)}_{\lambda,j}.
\notag
\end{align}
At this stage, the only approximation that was used concerns the truncation of the Fourier space into $2n_F\!+\!1$ modes. In particular, no assumption was made on the frequency of the shaking, which implies that Eq.~\eqref{FourierElement} is valid beyond the rotating-wave approximation (i.e.~in the regimes of slow time-modulation). However, in the following we set the system parameters within the range of validity of the rotating-wave approximation, which allows for a good comparison between the results obtained from the full time-dependent Hamiltonian in Eq.~\eqref{FourierElement} and those stemming from the effective model defined in Sec.~\ref{Sec:ShakenChannel}.
Moreover, we point out that Eq.~\eqref{FourierElement} explicitly involves the energy $\omega \lambda$, which indicates that the energetic nature of the synthetic dimension is implicitly present in the description. 

We now apply the methods described in Appendix~\ref{Sec:TransportFloquet} to numerically compute the conductance of the system described by Eq.~\eqref{FourierElement}, by generalizing Eq.~\eqref{zeroConductance} for time-periodic systems. Due to the driven nature of the system, it is useful to define a ``Fermi quasi-energy" $\varepsilon_F \in [-\omega/2, \omega/2]$ such that $E_F = \varepsilon_F + n\omega$, where $E_F$ refers to the Fermi energy set by the unshaken reservoirs. As discussed in Refs.~\cite{Farrell, Yap}, the conductance of the effective model is then recovered by summing the transmissions $T_{\alpha,\beta}^{(m)}$ attributed to the different Floquet multiplicities 
\begin{align}
 T_{\alpha,\beta}^{(m)}(\varepsilon_F)= \sum_{n \in \mathbb{Z}} T_{\alpha,\beta}^{(m)}(\varepsilon_F + n \omega), 
 \label{FloquetSumRule}
 \end{align}
which is then to be combined with Eq.~\eqref{zeroConductance}. 
The Floquet sum rule in Eq.~\eqref{FloquetSumRule} allows one to recover the conductance of the effective (Floquet) time-independent model~\cite{Yap}.

Specifically, the conductance is calculated by considering a two-terminal geometry, namely, by explicitly using the fact that the shaken channel is physically connected to two reservoirs [Fig.~\ref{Fig:TransportChannel}]. As stated above, we stress that the energetic nature of the synthetic dimension is naturally included in the formalism through Eq.~\eqref{FourierElement}. The next paragraphs demonstrate how this full-time-dependent approach reproduces the features that were previously obtained based on the effective-Hamiltonian (multi-terminal) approach described in Sec.~\ref{Sec:EffMultiTerminal}. 

\subsection{Time-dependent model with homogeneous hopping} 
\label{Sec:FloquetIsotropy}

As a first step, we propose to analyze the transport properties of the full time-dependent model by supposing that all the hopping processes in Eq.~\eqref{FourierElement} are uniform and isotropic over the entire (fictitious) 2D lattice [see Sec.~\ref{Sec:EffNoAnisotropy}]. In this way, one will be able to directly compare the results obtained from the Floquet-Landauer approach with those previously presented in Sec.~\ref{Sec:Eff2Terminal} (where the effective-Hamiltonian approach was studied based on a simple two-terminal geometry); in particular this will shed some light on the Floquet sum rule [Eq.~\eqref{FloquetSumRule}], in the context of a time-dependent two-terminal setup. 

The results presented in this Section have been obtained using the Floquet RGF method described in Appendix~\ref{Sec:TransportFloquet}; in the present model, numerical convergence is reached for $n_F \!=\! \lambda_\text{cut}$, where $\lambda_\text{cut}$ denotes the cut-off along the synthetic dimension.

\begin{figure}[t]
\centering
\includegraphics[width=1 \columnwidth]{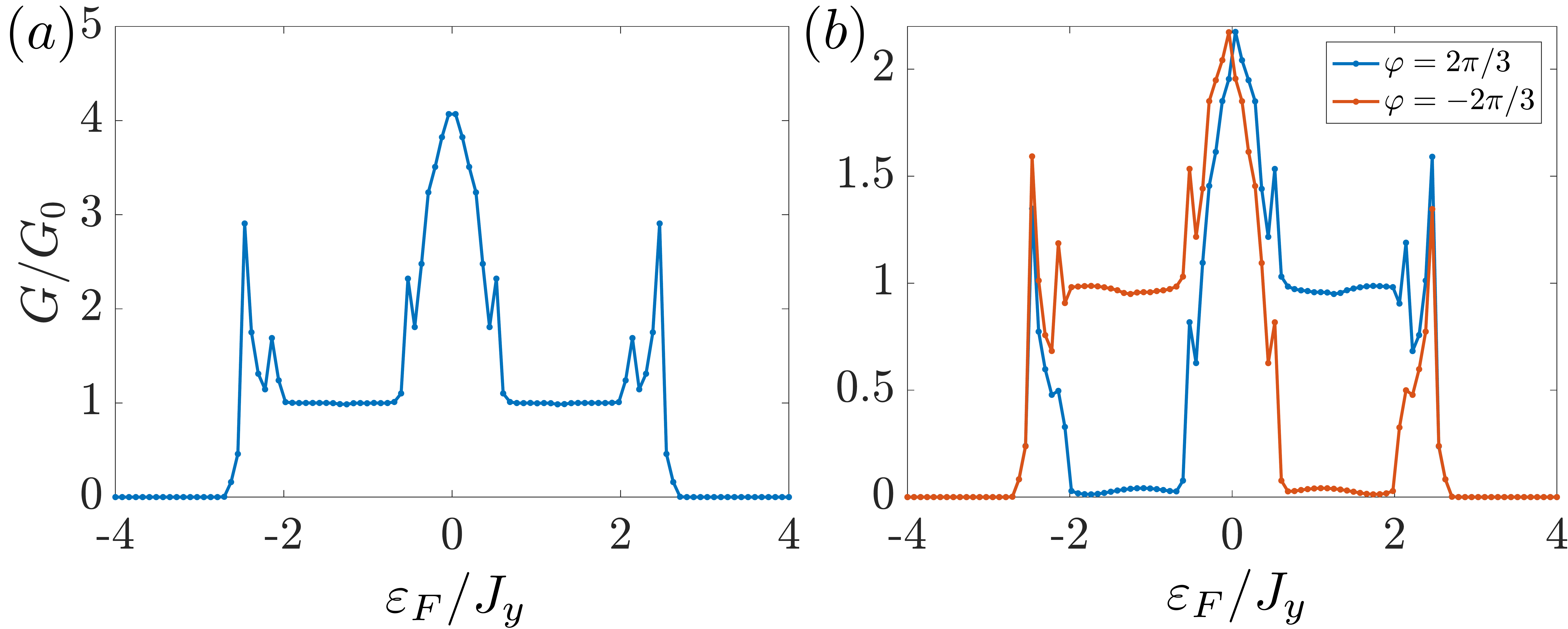}
\caption{Conductance of the shaken-channel model described by Eq.~\eqref{Schroedinger}, when neglecting the anisotropy in the hopping, as obtained from the Floquet RGF method and Floquet sum rule in Eq.~\eqref{FloquetSumRule}. Parameters are $\lambda_\text{cut}=11$, $N_y=12$, $\varphi=2\pi/3$, $J_y=J_1$, $\kappa=0.03\omega/l_H$, and $n_F\!=\!11$. (a) The sum in Eq.~\eqref{FloquetSumRule} is taken over all the Floquet modes, from $n\!=\!-11$ up to $n\!=\!n_F\!=\!11$; (b) the sum is performed up to the $n\!=\!5$ component.}
\label{Fig:Floquet}
\end{figure}

We illustrate the results in Fig.~\ref{Fig:Floquet}(a), which shows the conductance of the shaken-channel [Eq.~\eqref{Schroedinger}] with $N_y\!=\!12$ lattice sites along the real direction $y$, and we have set $\lambda_\text{cut}\!=\!11$. Here, we have applied the Floquet sum rule by including the contribution of all Fourier modes [Eq.~\eqref{trans_Floquet_sum}], which, as discussed in Ref.~\cite{Yap}, allows for an accurate evaluation of the conductance associated with the underlying effective (time-independent) system; in the present case, this leads to the clear quantized plateaus in Fig.~\ref{Fig:Floquet}(a), in agreement with the effective-Hamiltonian result in Fig.~\ref{Fig:RealisticCoupling}. 

As a technical remark, we note that the results are obtained within the validity range of the wide-band approximation [see Appendix~\ref{Sec:TransportMethods}]. In the context of time-periodic systems, this approximation should be extended by assuming that the self-energy and the linewidth are energy independent for all the Floquet multiplicities that contribute to the sum rule in Eq.~\eqref{FloquetSumRule}; as discussed in Ref.~\cite{Yap}, this is required to accurately capture the conductance.

\subsubsection*{Identification of chiral edge modes using the Floquet sum rule}

As illustrated above, the Floquet sum rule [Eq.~\eqref{FloquetSumRule}] allows one to evaluate the conductance associated with the underlying effective model~\cite{Yap, Farrell}. In order to test this result, we show in Fig.~\ref{Fig:Floquet}(b) the conductance as obtained by truncating the sum over the Fourier modes up to some critical mode $n_{\text{cut}}$:~by restricting this sum to $n\!=\!-n_F, \dots, 5$, we find that the plateau at $\varepsilon_F\!<\!0$ is drastically reduced, while the quantized plateau at $\varepsilon_F\!>\!0$ survives the truncation [compare the blue curve in Fig.~\ref{Fig:Floquet}(b) with Fig.~\ref{Fig:Floquet}(a)]. Besides, we find that this behavior is reversed (i.e.~only the plateau at $\varepsilon_F\!<\!0$ survives) upon reversing the sign of the magnetic flux ($\varphi\!=\!-2\pi/3$). This suggests an interesting interplay between the truncation of the Floquet sum rule, the energetic nature of the synthetic dimension and the detection of edge modes, as we now explain.

From our analysis of the effective Hamiltonian, we know that the quantized plateau at $\varepsilon_F\!>\!0$ [resp.~$\varepsilon_F\!<\!0$] is due to the edge mode localized at $\lambda\!=\!0$ [resp.~$\lambda\!=\!\lambda_\text{cut}$]. When the Floquet sum rule is truncated up to some mode $n_{\text{cut}}\!\ll\!\lambda_\text{cut}$, the total transmission $\sum_{n\! <\! n_{\text{cut}}} T(\varepsilon_F+n\omega)$ no longer captures the contribution of the edge mode localized at $\lambda\!=\!\lambda_\text{cut}$; this explains the absence of the expected plateau at $\varepsilon_F\!<\!0$ in Fig.~\ref{Fig:Floquet}(b) [blue curve]. In contrast, the quantized plateau at $\varepsilon_F\!>\!0$ is still present, since the total transmission still captures the contribution of the edge mode localized at $\lambda\!=\!0$. This observation is further validated by reversing the magnetic flux ($\varphi=-2\pi/3$): in this case, the propagating edge mode at $\lambda\!=\!0$ corresponds to the lower gap $\varepsilon\!<\!0$, and hence, it is the plateau at $\varepsilon_F\!>\!0$ that disappears [red curve in Fig.~\ref{Fig:Floquet}(b)].

This apparent relation between the Fourier modes ($n$) entering the Floquet sum rule and the ``site'' index $\lambda$ associated with the synthetic dimension naturally stems from the resonant nature of the time-modulation, which is at the heart of the present proposal. Indeed, the harmonic-oscillator levels (i.e.~the ``sites'' along the synthetic dimension $\lambda$) are equispaced according to the energy separation $\omega$ set by the trap frequency, which also corresponds to the separation between the many multiplicities ($n$) associated with the Floquet spectrum (since $\omega_D\!=\!\omega$); see Fig.~\ref{Fig:Scheme_system} in Appendix~\ref{Sec:Floquet_Landauer}. Furthermore, as we have discussed, the junction regions that connect the reservoirs to the inner system [Fig.~\ref{Fig:Junction}] can be thought of as uncoupled harmonic-oscillator levels, and this leads to an effective multi-terminal geometry where each row of ``sites'' corresponding to a given $\lambda$ is connected to two individual (fictitious) reservoirs ($L_{\lambda}$ on the left, and $R_{\lambda}$ on the right); see Sec.~\ref{Sec:EffMultiTerminal} and Fig.~\ref{Fig:TwoTerminalsAnistropy}(c). In this picture, calculating the contribution of the $n$-th mode to the conductance, $G(\varepsilon_F +n \omega)$, is related to selecting the effective ``terminals'' $L_{n}$ and $R_{n}$ that are energetically resonant with the mode, namely, the terminals that are connected to the ``sites'' $\lambda\!=\!n$. In this sense, scanning through the Fourier modes is reminiscent of analyzing various transport channels in the effective multi-terminal geometry [see Fig.~\ref{Fig:TwoTerminalsAnistropy}(c)]:~in particular, the contribution of the edge modes localized at $\lambda\!=\!0$ can be identified through the conductance $G_{L_0,R_0}$ associated with the lower ``terminals''. In practice, measuring the contribution of a given Fourier mode, $G_{L_n,R_n}$, can be achieved by setting the Fermi energy on resonance with respect to the corresponding harmonic oscillator states ($\lambda\!=\!n$); see also Appendix~\ref{Sec:Floquet_Landauer}. 

These observations lead to a remarkable corollary:~Restricting the Floquet sum rule to a limited number of modes (set by $n_{\text{cut}}$) can be used as a method to isolate the contribution of individual chiral edge modes. This particular feature of our synthetic-dimension system allows for the unambiguous detection of the quantized Hall conductance in a two-reservoir setting. We will further illustrate this important result in the next Section~\ref{Sec:FloquetFull}, based on the complete time-dependent model [Fig.~\ref{Fig:FloquetFull}].

As a technical remark, we note that the Floquet sum rule can be restricted to positive modes ($n\!\geq\!0$) in the present context, as the transmissions $T(\varepsilon_F+n\omega)$ associated with negative $n$'s are found to have negligible contributions.

\subsection{Full time-dependent model}
\label{Sec:FloquetFull}

\begin{figure}[t]
\centering
\includegraphics[width=1 \columnwidth]{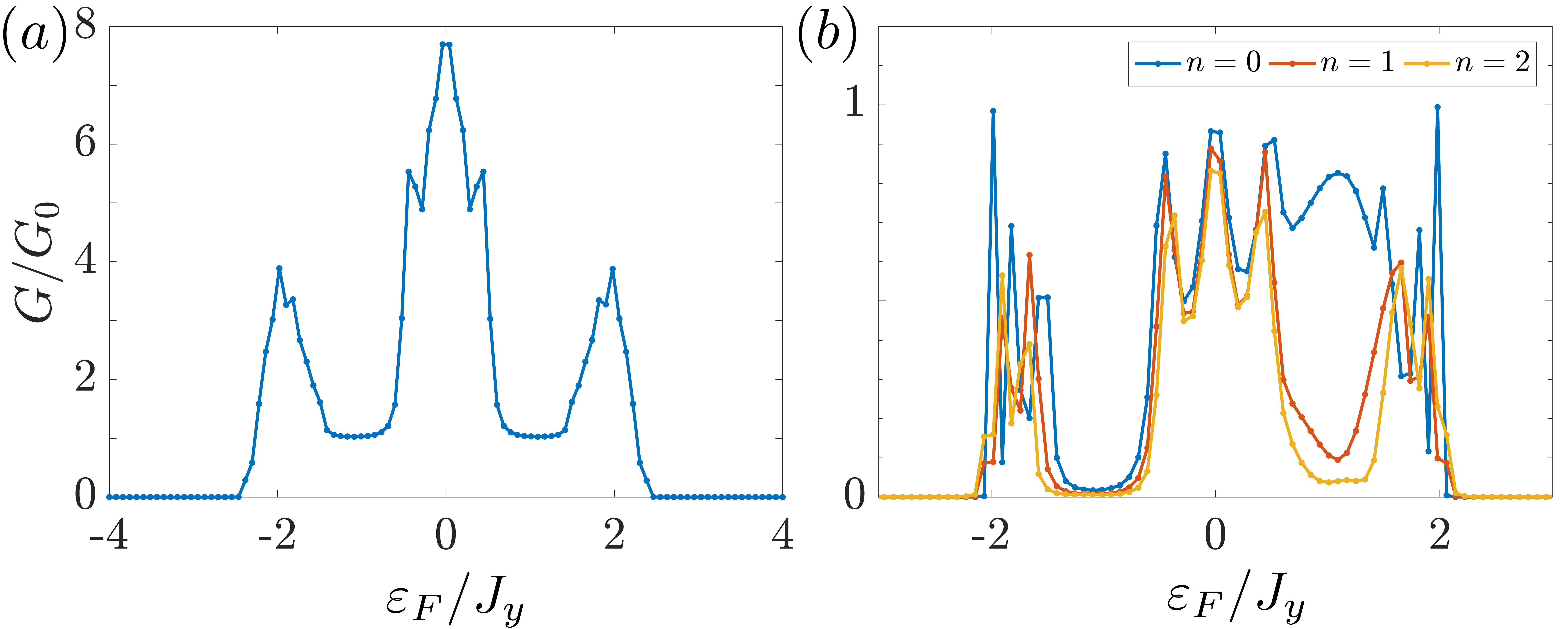}
\caption{Two-terminal conductance of the complete shaken-channel model defined in Eq.~\eqref{Schroedinger}, as obtained from the Floquet RGF method. Parameters are $\lambda_\text{cut}=11$, $N_y=12$, $\varphi=2\pi/3$, $J_y=4J_1$, $\kappa=0.03 \omega/l_H$, and $n_F\!=\!11$. (a) The conductance is obtained from the complete Floquet sum rule in Eq.~\eqref{FloquetSumRule}; (b) individual contributions to the conductance, corresponding to $n\!=\!0$ (blue), $n\!=\!1$ (orange) and $n\!=\!2$ (green). The main contribution to the plateau at $E_F\!>\!0$ stems from the $n\!=\!0$ component.}
\label{Fig:FloquetFull}
\end{figure}

We finally discuss the transport properties of the full time-dependent model in Eq.~\eqref{Schroedinger}, without neglecting the anisotropic hopping along the synthetic dimension $\lambda$. Here, we chose the hopping parameters $J_y\!=\!4J_1$, such that the band-gaps remain well open for $\lambda\!<\!30$; see Sec.~\ref{Sec:EffAnisotropy}.

We show the corresponding two-terminal conductance in Fig.~\ref{Fig:FloquetFull}(a), which was calculated using the complete Floquet sum rule (in the wide-band limit). This result reproduces the quantized plateaus of Fig.~\ref{Fig:TwoTerminalsAnistropy}, which was obtained using the effective-Hamiltonian approach. We note that the plateaus are slightly larger in Fig.~\ref{Fig:FloquetFull}(a), which is due to the fact that the band-gap is slightly larger in the present configuration where $\lambda_\text{cut}\!=\!11$, and $N_y\!=\!12$.

In Fig.~\ref{Fig:FloquetFull}(b), we show individual contributions to the total conductance $G(\varepsilon_F+n\omega)$, considering the first three Fourier modes $n\!=\!0,1,2$; these are the first three non-vanishing contributions to the Floquet sum rule used in Fig.~\ref{Fig:FloquetFull}(a). As previously discussed in Sec.~\ref{Sec:FloquetIsotropy}, the main contribution to the plateau at $E_F>0$ is due to the edge mode localized at $\lambda\!=\!0$, which is selected by the lowest $n\!\simeq\!0$ Fourier component entering the sum rule. As can be seen in Fig.~\ref{Fig:FloquetFull}(b), the contribution of the $n\!=\!1$ component to the edge-mode signal is already substantially reduced, which is due to the highly localized nature of the edge mode. Similarly, we have verified that the main contribution to the plateau at $\varepsilon_F<0$ mainly comes from the component $n_F\!=\! \lambda_\text{cut}$. 

The result shown in Fig.~\ref{Fig:FloquetFull}(b) demonstrates how the quantized conductance associated with a topological edge mode (here, at $\lambda\!=\!0$) can be unambiguously detected using a few conductance measurements, in a single atomic wire.

\section{Experimental considerations}
\label{Sec:Exp}
We now discuss the experimental implementation of the shaken-channel scheme, based on the demonstrated two-terminal cold-atom setup of Refs.~\cite{Brantut, Krinner_Nature}. As shown in Fig.~\ref{Fig:TransportChannel}, the quantum wire consists of a region with a tight harmonic confinement along both $x$- and $z$-directions. The propagation of atoms from one reservoir to the other in the $y$-direction is ballistic~\cite{Brantut, Krinner_Nature, Krinner_Rev}. We consider, in line with Refs.~\cite{Brantut, Krinner_Nature}, that the temperature is low enough such that individual harmonic-oscillator states in the wire can be resolved by transport, yielding a quantized conductance upon varying the chemical potential [see Section~\ref{Sec:FloquetFull}]. 

Using standard high-resolution optical techniques, the wire can be exposed to a periodic drive, while keeping the adiabatic connection to non-shaken reservoirs.
The position-dependent phase of the temporal modulation in Eq.~\eqref{timemod} can be realized using Raman transitions between harmonic-oscillator states along one transverse direction. Alternatively, a direct time- and space-periodic deformation of the wire structure could be engineered based on light shaping techniques. 
The typical length that can be achieved in these quantum wires is from ten to twenty micrometers. 
The spatial period $l_B \!=\! 2\pi/\varphi$ associated with the time-dependent potential in Eq.~\eqref{timemod} must be much shorter than the length of the wire itself. 
The driving strength $\kappa$ is controlled by the intensity of the Raman beams or the amplitude of the deformation and should satisfy the two following criteria: (i) ensure the validity of the rotating-wave approximation~\cite{Price} for the relevant (low-energy) harmonic-oscillator states participating to transport, which places an upper bound on the driving strength $\kappa$; (ii) create a topological bulk gap (hosting the edge modes) larger than temperature in order to be resolved by transport measurements, which constrains the strength from below; 
based on the effective band-gap of Fig.~\ref{Fig:TwoTerminalsAnistropy} (a), we estimate the band-gap to be  $J_y\!\approx\!2\kappa\sqrt{1/2 M \omega}$, which has to fulfill $k_{\text{B}} T_{\text{temp}}\!\leq\! J_y$, with $T_{\text{temp}}$ the temperature of the atoms in the reservoirs.
While these are required for the above formalism to apply, we expect the physics to be robust against moderate deviations from these bounds. Besides, we note that realistic setups would typically involve 10-100 harmonic-oscillator states in the channel, hence offering a rather long synthetic dimension (as compared, for instance, to atomic-internal-states realizations~\cite{Mancini, Stuhl}), and therefore, a good resolution of the edge-state signal.

As previously noted, there is no need for the projection of a lattice structure along the transport direction for the observation of the chiral edge states in the synthetic dimension, even though such a projection has recently been demonstrated~\cite{Lebrat}. Indeed, without a lattice the model maps onto the coupled-wires model of Refs.~\cite{Kane, Budich}, known to exhibit the quantum Hall effect.

The natural observable in the experiment is the two-terminal conductance, measured as a function of the chemical potential. By repeating measurements for chemical potentials increased by an integer multiple of $\omega$, one can reconstruct the full conductance spectrum of the inner system, as indicated by the Floquet sum rule. 
A different type of measurement could be performed by making use of the direct observation of energy currents in two terminal systems, as demonstrated in Refs.~\cite{Brantut2013, Husmann2018}. Indeed, even without resolving the topological band structure, a chemical potential bias between the reservoirs will yield a current in the $\lambda$ direction, which will contribute to the thermopower of the channel and provide a direct measure of the chirality of the underlying model.

\section{Conclusions}
\label{Sec:Conclusions}

We have proposed a scheme by which the quantum Hall conductance of a neutral atomic gas can be detected using a minimal one-dimensional setting:~a quantum wire connected to two reservoirs~\cite{Brantut, Krinner_Nature}. The two-dimensional nature of the quantum Hall effect is offered by an additional (synthetic) dimension, which is naturally present in the system. Inspired by  Ref.~\cite{Price}, we proposed that a Chern-insulating state (realizing the quantum Hall effect) can be realized in this setting upon subjecting the quantum wire to a resonant modulation. Importantly, the resulting quantized Hall conductance can be unambiguously detected in this scheme, by exploiting an unusual feature offered by the synthetic dimension:~its energetic nature effectively leads to a multi-terminal geometry, which allows for a clear measurement of the chiral edge modes' contribution to transport. This appealing result was demonstrated using two complementary approaches, one based on effective (time-independent) Hamiltonians and the other on a Floquet-Landauer approach, which takes the full time-dependence of the problem into account.

Intriguing perspectives include the study of inter-particle interactions in this synthetic-dimension approach. As discussed in Ref.~\cite{Price}, interactions are long-ranged (but not infinite-range) along the synthetic dimension, and the corresponding phases are still to be elucidated. In particular, it would be interesting to identify regimes where strongly-correlated states with topological features could be stabilized in this setting; the corresponding Hall conductance could then be explored using the schemes and concepts detailed in the present work.   

The notions and results introduced in this work could be applied to other physical platforms. For instance, synthetic dimensions have been proposed in the context of photonics~\cite{Schmidt, Ozawa2016, Yuan, Ozawa2017, Lin2018}, and a first experimental realization -- reminiscent of the scheme proposed in Ref.~\cite{Price} -- was recently reported in Ref.~\cite{Lustig}. In this context, we note that a transport formalism (analogous to the Landauer formalism) has been proposed to describe transport of photons~\cite{Wang2016}. Altogether, this suggests that the scheme discussed in this work could be directly transposed to the context of topological photonics~\cite{Ozawa_Rev}. 

Finally, this work emphasized the richness and complexity of quantum transport in the general context of Floquet-engineered systems. As was highlighted in our study (and as particularly emphasized in Appendix~\ref{Sec:Floquet_Landauer}), a conventional configuration of the reservoirs (which creates a weak chemical-potential imbalance on either side of the system under scrutiny) leads to a intricate (non-uniform) occupation of the Floquet eigenstates associated with the inner system. While this generically complicated the analysis of such driven settings, we have shown how to take advantage of this unusual feature in order to finely probe and identify the transport properties of topological edge modes. We believe that such interplay between quantum transport, Floquet engineering and topological edge modes will play a crucial role in near-future experiments.

\begin{acknowledgments}
We acknowledge the Kwant code \cite{Groth}, which was used to benchmark some of the results shown in this paper.
We are grateful to Tomoki Ozawa, Alexandre Dauphin and Marco Di Liberto for fruitful discussions, and Philipp Fabritius for careful reading of the manuscript.
Work in Brussels was supported by the Fonds De La Recherche Scientifique (FRS-FNRS) (Belgium) and the ERC Starting Grant TopoCold.
H.M.P. is supported by the Royal Society via grants UF160112, RGF\textbackslash EA\textbackslash 180121 and RGF \textbackslash R1\textbackslash 180071
J.-P.B. is supported by the ERC project DECCA (project n$^\circ$714309) and the Sandoz Family Foundation-Monique de Meuron program for Academic Promotion.
L.C. is supported by ETH Zurich Postdoctoral Fellowship and Marie Curie Actions for People COFUND program and ERC Marie Curie TopSpiD (project n$^\circ$746150). M.L., S.H. and T.E.'s work is supported by ERC advanced grant TransQ (project n$^\circ$742579).
\end{acknowledgments}

\appendix
\section{Landauer-B\"{u}ttiker formalism and non-equilibrium Green's function}
\label{Sec:TransportMethods}
In this Appendix we review the theoretical framework offered by the Landauer-B\"{u}ttiker formalism, which was originally developed for calculating the conductance in solid-state systems~\cite{Landauer, Buttiker, Imry, Datta}, but which has also been recently applied to describe transport in charge-neutral atomic systems~\cite{Krinner_Rev}.
We first consider the case of a single channel connected to two external reservoirs, which act as contact terminals and are labelled by $\alpha$ and $\beta$; throughout, we will assume that particles obey Fermi statistics. The chemical potential in the $\alpha$ [resp.~$\beta$] reservoir is denoted $\mu_{\alpha}$ [resp.~$\mu_{\beta}$]; we will assume that these chemical potentials are centered around the Fermi energy $E_F$ and that their differences are small.
In this case, and assuming zero temperature for now, the linear d.c.~current that flows between the two reservoirs can be expressed as
\begin{equation}
I_{\alpha,\beta}= G_{\alpha,\beta}(E_F)  \Big(\mu_\alpha -\mu_\beta \Big),
\label{Landauer}
\end{equation}
where $G_{\alpha,\beta}(E_F)$ is the conductance of the system at a given Fermi energy $E_F$ \cite{Imry}.
This conductance can be evaluated using the Landauer formula
\begin{equation}
G_{\alpha,\beta}(E_F) \equiv \frac{1}{h}  \sum_m T_{\alpha,\beta}^{(m)}(E_F),
\label{zeroConductanceAppendix}
\end{equation}
which involves a sum over the $m$ possible transport channels of the transmission probabilities $T_{\alpha,\beta}^{(m)}(E_F)$ for a particle of charge $e$ to be carried through the system. The transmission probabilities are related to the scattering properties of the system \cite{Landauer}; in particular, if the system has $N$ perfectly transmitting channels, i.e. $T^{(m)}=1$ for all $m$, each of these will contribute with a quantum of conductance $G_0= 1/h$ such that the total conductance is quantized according to $G_{\alpha,\beta} = N G_0$.

For a system connected to many reservoirs, the d.c.~current at a given terminal $\alpha$ is generalized by the Landauer-B\"{u}ttiker formula
\begin{equation}
I_{\alpha}= \sum_{\beta} I_{\alpha, \beta} = \sum_{\beta} G_{\alpha,\beta}(E_F)  \Big(\mu_\alpha -\mu_\beta \Big), 
\label{Buttiker}
\end{equation}
where the sum is now taken over all other reservoirs $\beta$, which are connected to $\alpha$.

In the case of finite temperature, the currents in Eq.~\eqref{Landauer} and Eq.~\eqref{Buttiker} must be weighted with the Fermi-Dirac distributions evaluated at the two reservoirs $f_{\alpha}\equiv f(E-	\mu_{\alpha})$ and $f_{\beta}\equiv f(E-	\mu_{\beta})$, and then integrated over all energies, which yields the following generalized expression \cite{BruusFlensberg}:
\begin{equation}
I_{\alpha} \equiv \frac{1}{h}\sum_{\beta,m} \int  T_{\alpha,\beta}^{(m)}(E) \left[f_\alpha-f_\beta\right] \mathrm{d}E.
\end{equation}

To evaluate the transmission probabilities in Eq.~\eqref{zeroConductanceAppendix}, it is often convenient to use the non-equilibrium Green's function method, which is mathematically equivalent to the scattering approach in the linear regime \cite{Datta, Ryndyk}. 
The matrix representation of the retarded Green's function $\boldsymbol{\mathcal{G}}^r$ of a system at Fermi energy $E_F$ is defined through the Hamiltonian matrix $\mathbf{H}$ as
\begin{equation}
\left[\mathbf{E} - \mathbf{H} \right]\boldsymbol{\mathcal{G}}^r= \mathbb{I},
\label{defGreen}
\end{equation} 
where $\mathbf{E}=\left(E_F+\mathrm{i}0^+\right)\mathbb{I}$, $\mathbb{I}$ is the identity matrix and $0^+$ is an infinitesimally small positive quantity. 

Without loss of generality, we focus on a channel connected to two reservoirs, which are attached to the left and the right of the system (in this setting, $\alpha, \beta= R, L$ refer to the two reservoirs).
The Hamiltonian matrix of the entire scattering system, including the reservoirs, has the following block structure
\begin{equation}
 \mathbf{H}_\text{tot}=\begin{pmatrix}
\mathbf{H}_L & \mathbf{H}_{LS} & 0\\
 \mathbf{H}_{LS}^\dagger & \mathbf{H}_S & \mathbf{H}_{RS}\\
0& \mathbf{H}_{RS}^\dagger & \mathbf{H}_R
 \end{pmatrix},
 \end{equation} 
where $\mathbf{H}_{L,R}$ refers to the Hamiltonians describing the left/right reservoirs, $\mathbf{H}_{S}$ describes the inner system (the transport channel), and where $\mathbf{H}_{LS}$ and $\mathbf{H}_{RS}$ describe the couplings between the inner system and the left/right reservoirs. Since the size of the reservoirs is typically very large, the size of the Hamiltonian matrices $\mathbf{H}_{L,R}$ is large compared to the size of $\mathbf{H}_S$.
From Eq.~\eqref{defGreen}, the Green's function of the total system is
\begin{equation}
\!\!
\begin{pmatrix}
\mathbf{E}-  \mathbf{H}_L & -\mathbf{H}_{LS} & 0\\
 -\mathbf{H}_{LS}^\dagger & \mathbf{E}- \mathbf{H}_S & -\mathbf{H}_{RS}\\
0& -\mathbf{H}_{RS}^\dagger &\mathbf{E}-\mathbf{H}_R
\end{pmatrix}^{\!\!-1}\!\!
\!=\! \begin{pmatrix}
 \boldsymbol{\mathcal{G}}^r_L & \boldsymbol{\mathcal{G}}^r_{LS} & 0\\
 {\boldsymbol{\mathcal{G}}^r_{LS}}^\dagger & \boldsymbol{\mathcal{G}}^r_S & \boldsymbol{\mathcal{G}}^r_{RS}\\
0& {\boldsymbol{\mathcal{G}}^r_{RS}}^\dagger & \boldsymbol{\mathcal{G}}^r_R
 \end{pmatrix}\!.
 \label{defGreenS}
\end{equation}
From Eq.~\eqref{defGreenS}, one can obtain the following relation for the Green's function of the inner system $\boldsymbol{\mathcal{G}}^r_S$
\begin{equation}
\left[\mathbf{E} - \mathbf{H}_S - (\boldsymbol{\Sigma}_L(E_F)+ \boldsymbol{\Sigma}_R(E_F)) \right]\boldsymbol{\mathcal{G}}^r_S= \mathbb{I}, 
\label{defGreen2}
\end{equation}
where $\boldsymbol{\Sigma}_L(E_F)\!=\!  \mathbf{H}_{L S}^\dagger (\mathbf{E}- \mathbf{H}_L)^{-1} \mathbf{H}_{L S}$ and $\boldsymbol{\Sigma}_R(E_F)\!=\!  \mathbf{H}_{R S} (\mathbf{E}- \mathbf{H}_R)^{-1} \mathbf{H}_{R S}^\dagger$. 
Equation \eqref{defGreen2} is very similar to Eq.~\eqref{defGreen}, except that the Hamiltonian is now modified with the term $\boldsymbol{\Sigma}(E_F)\!=\!\sum_{\alpha}\boldsymbol{\Sigma}_{\alpha}(E_F)$, the so-called contact self-energy, which includes the details of the reservoirs.
The anti-Hermitian counterpart of the self-energy $\boldsymbol{\Gamma}^\alpha(E_F) = \mathrm{i} \left(\boldsymbol{\Sigma}_\alpha(E_F) - \boldsymbol{\Sigma}^\dagger_\alpha(E_F)\right)$ defines the linewidth of the $\alpha$ reservoir, and it reflects the fact that particles are lost from the inner system due to leakage into the reservoirs; in this sense, the channel is out of equilibrium~\cite{Datta}.
The linewidth of the reservoir and the self-energy contributions can be obtained following standard prescriptions~\cite{Datta, Lewenkopf}.

Using the non-equilibrium Keldysh formalism \cite{Meir}, the transmission can be calculated through the Caroli formula \cite{Caroli}, which involves the Green's functions of the inner system and the self-energies of the reservoirs:
\begin{equation}
T_{\alpha,\beta}(E_F)=\mathrm{Tr}\left[\boldsymbol\Gamma^\alpha(E_F) \boldsymbol{\mathcal{G}}^r(E_F) \boldsymbol{\Gamma}^\beta(E_F) \boldsymbol{\mathcal{G}}^a(E_F)\right],
\label{Caroli}
\end{equation}
where we have omitted the subscript $S$ associated with the Green's functions of the system, for simplicity of notation; we note that the advanced Green's function satisfies $\boldsymbol{\mathcal{G}}^a = \left(\boldsymbol{\mathcal{G}}^r\right)^\dagger$.
Equation~\eqref{Caroli} is very convenient for numerical evaluations of the d.c.~current that flows between two terminals, when combined with the recursive Green's function (RGF) method~\cite{ThoulessRGF, Lewenkopf} based on the Dyson equation. The RGF method can be generalized to multi-terminal systems~\cite{Thorgilsson}, which we will use for calculating the Hall conductance later in this article. 

\subsubsection*{The wide-band approximation}\label{Section_wide}
We have seen that all the details of the reservoirs are included in the self-energy matrix $\boldsymbol{\Sigma}(E_F)$, which is used both for obtaining the linewidth $\boldsymbol{\Gamma}(E_F)$ and the Green's function of the inner system connected to the terminals.
In order to calculate the self-energy $\boldsymbol{\Sigma}(E_F)$, the reservoir is typically assumed to have a large volume and a high density of states. 
If the density of states of the reservoir is constant over an energy range much larger that the bandwidth of the inner system, the wide-band approximation can be used \cite{Haug}. 
Under this approximation, both the self-energy and the linewidth are taken to be energy independent. In particular $\boldsymbol{\Sigma}_\alpha \propto -\mathrm{i} \mathbb{I} \delta_{x,x_\alpha} \gamma$, where $\gamma$ is a constant that is of the order of the bandwidth of the inner system, and the $\delta_{x,x_\alpha}$ selects only the sites of the system that belong to the terminal $\alpha$ \cite{Yap}.
Unless otherwise stated, the wide-band approximation is used for all the results presented in the main text.

\section{Evaluating the conductance in periodically-driven systems: the Floquet-Landauer approach}
\label{Sec:TransportFloquet}

As we have seen, our proposal builds on the possibility of addressing a synthetic dimension by applying a time-periodic modulation to an atomic channel~\cite{Price}. In fact, this proposed scheme belongs to the general class of Floquet-engineered quantum systems, which aims at realize intriguing Hamiltonian models through periodic driving~\cite{Eckardt, Lignier, Kierig, Struck, GoldmanPRX, EckardtRMP, Esin}. In this context, it is common to derive an effective (Floquet) Hamiltonian that describes the long-time dynamics of the system, and which results from a rich interplay between the time-modulation and the underlying static system~\cite{GoldmanPRX, EckardtRMP}. 

In fact, Floquet engineering can also be exploited to transfigure quantum transport properties, in the sense that applying a temporal modulation can greatly modify the transport channels of a quantum system. In this quantum-transport framework, where the time-modulated system is further connected to reservoirs, it is generally insufficient to simply apply the standard tools of quantum-transport theory to the effective (Floquet) Hamiltonian, which describes the inner system; such a naive approach was discussed in Sec.~\ref{Sec:EffectiveTransport}. Instead, a more rigorous approach consists in using a generalization of the non-equilibrium Green's function method~\cite{Kohler, Tsuji, Kitagawa, Yap}, which is specifically tailored to treat time-periodic systems, as we now review in this Appendix; this approach was applied in Sec.~\ref{Sec:Floquet}. 

Consider a time-dependent Hamiltonian $\mathbf{H}(t+T)\!=\!\mathbf{H}(t)$, where $T\!=\!2\pi/\omega$ is the period of the applied temporal modulation. 
The Fourier expansion of the Hamiltonian yields 
\begin{equation}
\mathbf{H}(t) = \sum_{n=-n_F}^{n_F} \mathbf{H}_n e^{-\mathrm{i} n \omega t},
\label{HFourier}
\end{equation} 
where we have truncated the series up to $N_F\!=\! 2n_F+1$ modes. In Floquet systems, the energy is only defined up to multiples of the driving frequency $E\!=\! \varepsilon + n \omega$ (hereafter, we set $\hbar\!=\!1$). This leads to the notion of quasi-energies $\varepsilon$, which can be chosen within the Brillouin zone $\varepsilon\!\in\![-\omega/2, \omega/2]$; see the review~\cite{EckardtRMP}. The so-called Floquet spectrum, which is defined in this restricted range, is then periodically repeated for each multiplicity (i.e.~around each $n \omega$). As discussed in Ref.~\cite{Eckardt2015}, it is convenient to treat such periodically-driven systems in an extended (Floquet) Hilbert space, of dimension $N_x N_y N_F$, which explicitly takes these multiplicities into account. In this framework, the Hamiltonian is replaced by a so-called ``quasienergy operator'', $\boldsymbol{Q}$, whose components $Q_{m,n}\!=\!\mathbf{H}_{m-n} \!+\! \delta_{m,n} m \omega \mathbb{I}$ act in the original Hilbert space; here $\mathbf{H}_{n}$ refers to the Fourier components introduced in Eq.~\eqref{HFourier} and $\mathbb{I}$ is the $N_x\!\times\!N_y$ identity matrix.

Following Refs.~\cite{Kitagawa, Yap}, one generalizes Eq.~\eqref{defGreen2} in view of defining a Floquet representation for the Green's function through the relation
\begin{equation}
\left[\mathbf{E} -\mathbf{Q} - (\boldsymbol{\Sigma}_F^{R}+ \boldsymbol{\Sigma}_F^L) \right]\boldsymbol{\mathcal{G}}^r_F= \mathcal{I},\label{Floquet_relation_Green}
\end{equation}
where all matrices are defined in the extended Floquet Hilbert space of dimension $N_x N_y N_F$, and where $\mathcal{I}$ is the identity matrix in this extended space. Specifically, $\mathbf{E}$ and $\boldsymbol{\Sigma}_F$ are diagonal matrices whose elements are 
\begin{align}
\mathbf{E}&\equiv \mathrm{Diag}\left[\left(E_F+\mathrm{i}0^+\right)\mathbb{I}, \dots, \left(E_F+\mathrm{i}0^+\right)\mathbb{I} \right], \\
\boldsymbol{\Sigma}_F^{\alpha}& \equiv \mathrm{Diag}\left[\boldsymbol{\Sigma}_{\alpha}(E_F -n_F \omega), \dots, \boldsymbol{\Sigma}_{\alpha}(E_F +n_F \omega)\right],
\end{align}
where $\alpha\!=\!L,R$ and $\boldsymbol{\Sigma}_{\alpha}(E_F)$ is defined below Eq.~\eqref{defGreen2}. Solving Eq.~\eqref{Floquet_relation_Green} yields the Floquet representation for the Green's function, $\boldsymbol{\mathcal{G}}^r_F$:
\begin{equation}
\boldsymbol{\mathcal{G}}^r_F \equiv 
\begin{pmatrix}
\boldsymbol{\mathcal{G}}^r_{-n_F, -n_F} & \dots & \boldsymbol{\mathcal{G}}^r_{-n_F,0} & \dots &\boldsymbol{\mathcal{G}}^r_{-n_F,n_F} \\
\vdots & \ddots & \vdots & \dots &\vdots\\
\boldsymbol{\mathcal{G}}^r_{0, -n_F} & \dots & \boldsymbol{\mathcal{G}}^r_{0,0} & \dots &\boldsymbol{\mathcal{G}}^r_{0,n_F} \\
\vdots & \dots & \vdots & \ddots &\vdots\\
\boldsymbol{\mathcal{G}}^r_{n_F, -n_F} & \dots & \boldsymbol{\mathcal{G}}^r_{n_F,0} & \dots &\boldsymbol{\mathcal{G}}^r_{n_F,n_F} \\
\end{pmatrix}.
\end{equation}
As in Appendix~\ref{Sec:TransportMethods}, the linewidth can be defined as $\boldsymbol{\Gamma}^\alpha\!=\!\mathrm{i}\left({\boldsymbol{\Sigma}_\alpha}_F - {\boldsymbol{\Sigma}_\alpha^\dagger}_F \right)$, and it can also be represented in the extended Hilbert space (with components denoted $\boldsymbol{\Gamma}^\alpha_{n,m}$). 

The Floquet generalization of the Caroli formula in Eq.~\eqref{Caroli} can then be obtained by treating the Fourier components of the Green's function and the linewidth individually. The resulting Floquet-Caroli formula for the transmission, at a given Fermi energy $E_F$, reads \cite{Yap}
\begin{align}
&T_{\alpha,\beta}(E_F)= \label{FloquetCaroli}\\
&\sum_{n =-n_F}^{n_F} \mathrm{Tr}\left[\boldsymbol{\Gamma}^\alpha_{0,0}(E_F) \boldsymbol{\mathcal{G}}^r_{n,0}(E_F) \boldsymbol{\Gamma}^\beta_{n,n}(E_F) \boldsymbol{\mathcal{G}}^a_{0,n}(E_F)\right],
\notag
\end{align}
where $\boldsymbol{\Gamma}^\alpha_{n,m}$ denote the components of the linewidth $\boldsymbol{\Gamma}^\alpha$ introduced above.

\subsection*{Floquet sum rule}
\label{Sec:FloquetSumRule}
In the framework of periodically-driven systems, one is interested in studying the contribution of quasi-energy bands $\varepsilon(\boldsymbol{k})$ to the conductance. In particular, if a quasi-energy band is associated with a non-zero Chern number, one would expect to observe a quantized Hall conductance, in direct analogy with the static case. However, this analysis requires special care, as previously highlighted in Refs.~\cite{Farrell, Yap,Kitagawa}. 

In order to draw an analogy with static systems, one introduces a Fermi quasi-energy $\varepsilon_F $, which is defined as $E_F\!=\!\varepsilon_F + n \omega$, where $E_F$ is the Fermi energy set by the static reservoirs, and where $\varepsilon_F \in [-\omega/2, \omega/2]$ scans the quasi-energy spectrum within a single Floquet-Brillouin zone. Interpreting the quasi-energy spectrum as the energy spectrum associated with a static system, one is interested in evaluating the conductance at a given $\varepsilon_F $ (i.e.~interpreting $\varepsilon_F $ as the standard Fermi energy). As previously shown in Refs.~\cite{Farrell, Yap}, this conductance calculation can be achieved by summing over the transmissions associated with all multiplicities~\cite{Farrell, Yap}:
\begin{equation}
 T_{\alpha,\beta}(\varepsilon_F)= \sum_{n \in \mathbb{Z}} T_{\alpha,\beta}(\varepsilon_F + n \omega).
 \label{FloquetSumRuleAppendix}
 \end{equation} 
As illustrated in the main text, the sum rule in Eq.~\eqref{FloquetSumRuleAppendix} is essential to recover a quantized Hall conductance in periodically-driven systems realizing the quantum Hall effect (and Floquet Chern insulators in general~\cite{Kitagawa}). The transmission is calculated from Eq.~\eqref{FloquetCaroli} together with Eq.~\eqref{FloquetSumRuleAppendix}  as
\begin{align} 
&T_{\alpha,\beta}(\varepsilon_F)=\!\sum_{n,n^\prime= -n_F, \dots, n_F}\!\mathrm{Tr}\Big[\boldsymbol{\Gamma}^\alpha_{0,0}(\varepsilon_F+n\omega)  \label{trans_Floquet_sum} \\ 
&\times \boldsymbol{\mathcal{G}}^r_{n^\prime,0}(\varepsilon_F+n\omega)  \boldsymbol{\Gamma}^\beta_{n^\prime,n^\prime}(\varepsilon_F+n\omega) \boldsymbol{\mathcal{G}}^a_{0,n^\prime}(\varepsilon_F+n\omega)\Big] \notag.
\end{align}
 
The sum rule in Eq.~\eqref{FloquetSumRuleAppendix} also leads to an interesting experimental corollary, which is that the conductance cannot be evaluated based on a single measurement. As discussed in Ref.~\cite{Yap}, the transport experiment should be repeated for various values of the reservoirs' chemical potential, which should be chosen so as to probe the many multiplicities $\varepsilon_F + n \omega$. The convergence of this approach is illustrated in Section~\ref{Sec:Floquet}. 

\section{The Floquet eigenstates in the channel and the bare levels in the static reservoirs}
\label{Sec:Floquet_Landauer}

\begin{figure}[t!]
\centering
\includegraphics[width= 1 \columnwidth]{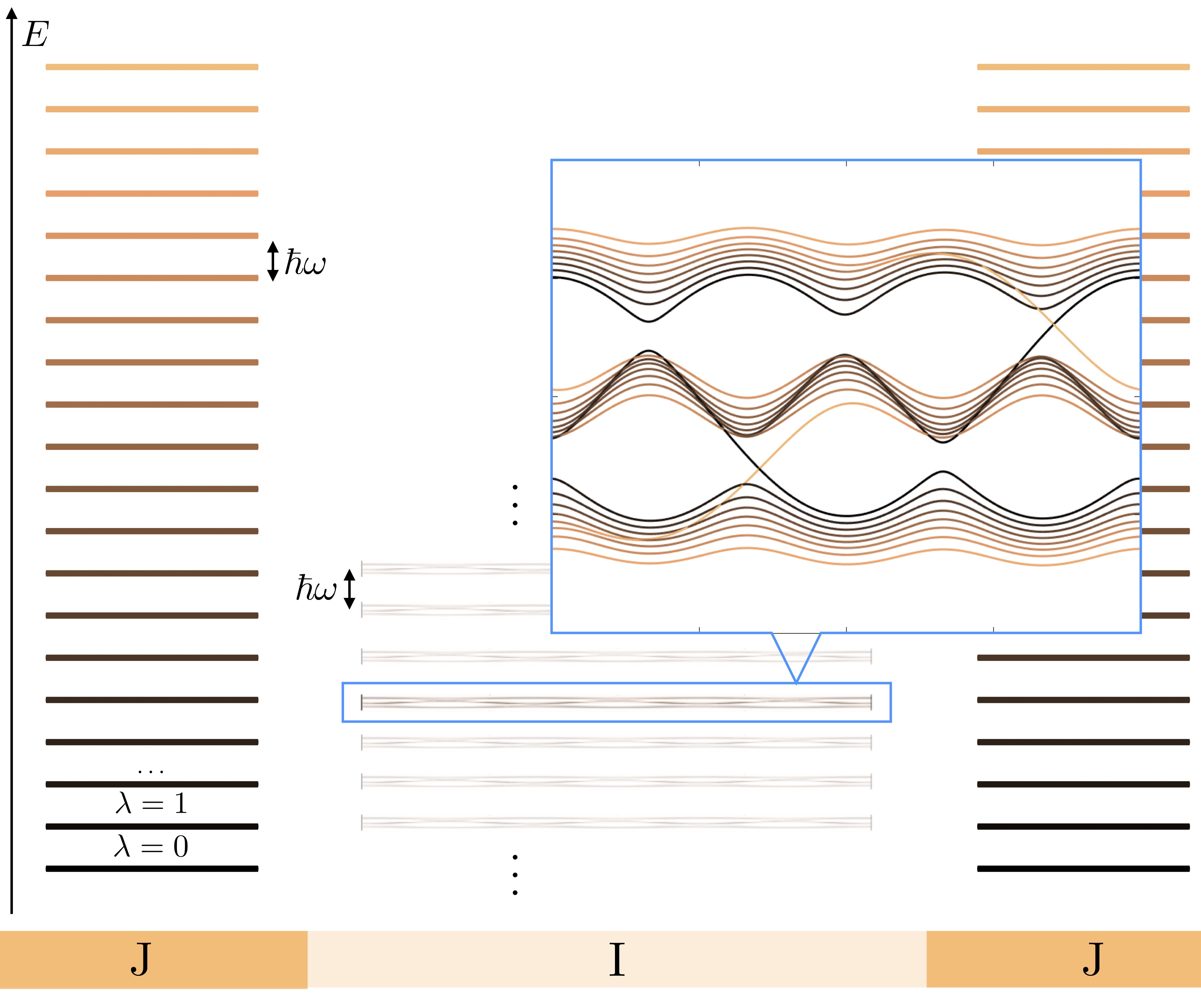}
\caption{Schematic representation of the energy levels of the junction regions (J) and the inner region (I). The color scale is chosen as in Fig.~\eqref{Fig:TwoTerminalsAnistropy}(a), and indicates the mean position $\langle \lambda \rangle$ of the states. The levels in the junction region are the harmonic oscillator states, separated by $\lambda \omega$. The quasi-energy spectrum $\varepsilon (k_y)$ of the inner region contains contributions from all the $\lambda$ states, due to the coupling provided by the shaking. Here the Floquet spectrum is represented in an extended zone scheme, according to which the same Floquet eigenstates are repeated periodically, $E = \varepsilon + n\omega$. We note that the separation between these Floquet multiplicities matches the separation between the bare $\lambda$ states. The dispersion of the states in the junction, due to propagation along the $y$ direction, is associated with a small bandwidth $J_y\!\ll\hbar \omega$, which is of the same order as the bandwidth of the quasi-energy spectrum in the channel. 
The population of the Floquet eigenstates within the channel is non-thermal, but reflects the thermal population in the undriven reservoirs (see text).}
\label{Fig:Scheme_system}
\end{figure}

This Appendix aims at further deepening our understanding of the Floquet system studied in Sec.~\ref{Sec:Floquet}, by analyzing how the states of the static system (i.e.~the reservoir and junction regions) project onto the states of the shaken system (inner region). 

We start by displaying in Fig.~\ref{Fig:Scheme_system} a schematic picture of the energy levels in the channel, where the color scale highlights the low-$\lambda$ levels with darker colors. 
In the junction regions, the energy levels are the equally-spaced harmonic oscillator states, labelled by $\lambda$.
Within the inner region, the driving protocol couples all the $\lambda$ states together, and the quasi-energy dispersion $\varepsilon$ corresponds to that of the effective model in Eq.~\eqref{Heff}, whose bandwidth is of order $J_y$. Due to the driven nature of the inner system, this quasi-energy spectrum can be  repeated periodically, $E\!=\!\varepsilon + n\omega$, where $n$ labels the many replicates. As indicated in Fig.~\ref{Fig:Scheme_system}, the spacing between these multiplicities also corresponds to the energy separation between the bare $\lambda$ states in the junction regions; it should also be noted that the states in the junction regions are associated with a finite dispersion due to the motion along the $y$ direction, which is characterized by a bandwidth of order $J_y$.

We now address the following question: Considering that the Fermi energy $E_F$ set in the reservoirs is such that only the first few low-$\lambda$ states are populated, what are the (Floquet) eigenstates of the inner region that are predominantly occupied and hence contribute to transport? 
Importantly, we stress that the population of the different energy levels within the channel is non thermal, but reflects the thermal population of the uncoupled $\lambda$ states in the reservoirs.
Although very schematic, we already observe from Fig.~\ref{Fig:Scheme_system} that the low-$\lambda$ states in the junction regions (black) mainly overlap with the mid-gap edge states of the inner (shaken) region, with only small overlaps with bulk states; hence, we expect these edge states to be significantly populated in this reservoir configuration. 

\begin{figure}[t!]
\centering
\includegraphics[width=1 \columnwidth]{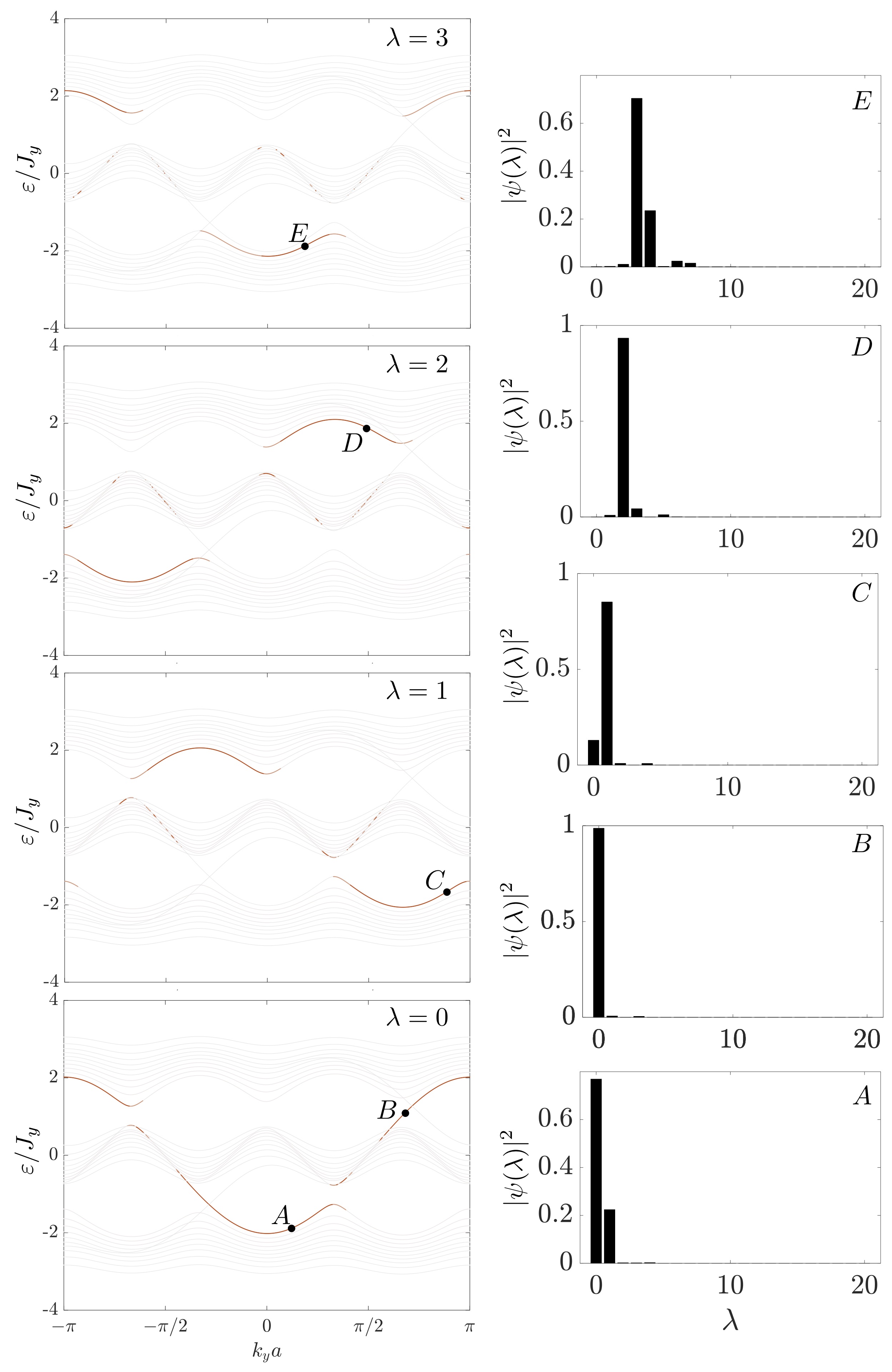}
\caption{Energy spectra as in Fig.~\ref{Fig:TwoTerminalsAnistropy}(a), where we only show the parts of the spectra whose states are localized around $\lambda = 0,1,2,3$ respectively. The amplitudes $|\psi(\lambda)|^2$ of the states indicated with a dot are also shown.}
\label{Fig:Fig7_decomposed}
\end{figure}

To further illustrate this point, we plot in Fig.~\ref{Fig:Fig7_decomposed} the energy spectrum of the effective time-independent model and highlight those states that are strongly localized around $\lambda=0,1,2,3$, respectively (we recall that this model is defined in the $\lambda-y$ plane, where $\lambda$ refers to the synthetic-dimension coordinate). We also plot the corresponding amplitudes $|\psi(\lambda)|^2$ for the special states indicated by black dots. 
We note that the mid-gap edge state (i.e.~the $B$ state in Fig.~\ref{Fig:Fig7_decomposed}) is indeed sharply localized along $\lambda=0$, and that the bare states on the next row ($\lambda\!=\!1$) already have a very little overlap with it.
From this very simple ``decomposition'' of the inner-system quasi-energy spectrum [see also Fig.~\ref{Fig:TwoTerminalsAnistropy}(a) of the main text], we deduce that a chiral edge transport occurs along the channel (i.e.~along the $\lambda\!=\!0$ axis of the hybrid 2D system) whenever the Fermi energy is set in the vicinity of the lowest harmonic-oscillator state $\lambda$ in the reservoirs. This is compatible with the result shown in Fig.~\ref{Fig:FloquetFull}(b), where the main contribution to the plateau in the conductance spectrum (at positive energy) was shown to result from the $n\!=\!0$ Fourier component, namely, when the Fermi energy $E_F\!=\!\varepsilon_F + n \omega$ was set around the $\lambda\!=\!0$ state of the harmonic oscillator.  \\

It is important to notice that although the reservoir imposes a standard (thermal) Fermi distribution to the junction region, the bare $\lambda$ states project very differently on the various states of the Floquet inner system, hence giving rise to a non-standard (non-thermal) distribution of populations. In fact, the state occupation depends also on the localization along the synthetic dimension, and that is a crucial difference with the usual Fermi distribution:~setting a given Fermi energy $E_F\!=\!\varepsilon_F + n \omega$ in the reservoir \textit{does not} imply that all the states in the inner system are empty if they lie above $\varepsilon_F$, or that they are occupied if they are below $\varepsilon_F$.

\end{document}